\documentclass[aps,onecolumn,prd,nofootinbib,superscriptaddress,preprintnumbers,longbibliography,eqsecnum,notitlepage,english,10pt]{revtex4-1}
\usepackage{amssymb,amsfonts,amsmath,blkarray,bm,dsfont,fixltx2e,nicefrac,mathtools,soul}
\usepackage{babel}
\usepackage{booktabs}
\usepackage{graphicx}
\usepackage{adjustbox}
\makeatletter
\let\old@makecaption=\@makecaption
\usepackage{caption,subcaption}
\let\@makecaption=\old@makecaption
\makeatother
\usepackage[dvipsnames]{xcolor}
\usepackage{multirow}
\usepackage{titletoc}
\usepackage{hyperref}
\hypersetup{
    colorlinks=true,
    urlcolor=blue,
    linkcolor = blue,
    urlcolor  = blue,
    citecolor = blue,
    anchorcolor = blue,
    pdftitle={a1(1420)},
    pdfauthor={Bonn-Bern-Washington}
}
\usepackage{cleveref} 
    \crefformat{figure}{Fig.~#2#1#3}
    \crefformat{table}{Tab.~#2#1#3}
    \crefformat{equation}{Eq.~(#2#1#3)}
    \crefformat{section}{Sect.~#2#1#3}
\graphicspath{{./figures/}}
\newcommand{\cg}[3]{(#1~#2|#3)}

\newcommand{\MeV}{{\rm ~MeV}}

\newcommand{\diff}{\mathrm{d}}
\usepackage{tikz}
\usepackage[compat=1.1.0]{tikz-feynman}
\tikzfeynmanset{every edge/.append style={very thick}}
\tikzset{
  cutline/.style={
    draw=orange,
    dashed,
    line width=4pt,
    dash pattern=on 10pt off 4pt   
  }
}
\usetikzlibrary{shadows}
\usetikzlibrary{shadows.blur}
\usetikzlibrary{backgrounds}
\usetikzlibrary{decorations.pathreplacing}
\tikzfeynmanset{
  triple/.style={
    draw,
    preaction={draw, line width=1pt,transform canvas={yshift=2pt}},
    preaction={draw, line width=1pt,transform canvas={yshift=-2pt}}
  }
}

\begin{document}
\title{\texorpdfstring{The $\bm{a_1(1420)}$ in a Unitary Coupled‑Channel Three‑Body Approach}{The a1(1420) in a Unitary Coupled‑Channel Three‑Body Approach}}
\preprint{JLAB-THY-26-4812}
\author{Ajay S. Sakthivasan}
\email{sakthivasan@hiskp.uni-bonn.de}
\affiliation{Helmholtz-Institut für Strahlen- und Kernphysik (Theorie) and Bethe Center for Theoretical Physics, Universität Bonn, 53115 Bonn, Germany}
\author{Yuchuan Feng}
\email{fengyuchuan@gwmail.gwu.edu}
\affiliation{The George Washington University, Washington, DC 20052, USA}
\author{Michael D\"oring}
\email{doring@gwu.edu}
\affiliation{The George Washington University, Washington, DC 20052, USA}
\affiliation{Theory Center, Thomas Jefferson National Accelerator Facility, Newport News, VA 23606, USA}
\author{Maxim Mai}
\email{maxim.mai@faculty.unibe.ch}
\affiliation{Albert Einstein Center for Fundamental Physics, Institute for Theoretical Physics, University of Bern, Sidlerstrasse 5, 3012 Bern, Switzerland}
\affiliation{The George Washington University, Washington, DC 20052, USA}

\begin{abstract}
An enhancement in the three-pion energy at around $\sqrt{s}\approx 1.42$~GeV with $a_1$ quantum numbers was observed at the COMPASS experiment. This was later attributed to the triangle singularity mechanism involving an on-shell $K^*(892)$, $K$ and $\bar K$ intermediate states. The alignment of the decay $K$ with the spectator $\bar K$ produces an $f_0(980)$, resulting in a kinematic enhancement, which is classically explained by the Landau equations. However, this one-loop process forms only part of a non-diagonal transition in a much larger coupled-channel framework. This study demonstrates the feasibility of embedding one-loop triangle-singularity calculations into a unitary three-body amplitude allowing one to consistently incorporate final-state interactions and their potentially substantial effect. For this, up to $P$-wave isobars and all sub-channel isospins are combined in a nine-channel production amplitude that is fitted to COMPASS lineshapes at different momentum transfers. The fitted amplitude reproduces the narrow enhancement in the $(\pi f_0)_P$ channel near $\sqrt{s}\approx1.42$ GeV. This implies that the triangle singularity mechanism sufficiently explains the observed enhancement, and an additional genuine $a_1(1420)$ pole is not required. Incidentally, the parameters of the ground state axial vector resonance (the $a_1(1260)$) are also extracted from that data.
\end{abstract}

\maketitle
\titlecontents{section}
  [2em]                              
  {\vspace{1pt}}                      
  {\contentslabel{1.5em}}             
  {}                                  
  {\titlerule*[0.5pc]{.}\contentspage} 
\titlecontents{subsection}
  [4em] 
  {\vspace{1pt}} 
  {\contentslabel{1.5em}} 
  {} 
  {\titlerule*[0.5pc]{.}\contentspage}
{\footnotesize
\tableofcontents\label{toc}
}
\setlength{\parskip}{4pt} 

\section{Introduction}
Quantum chromodynamics (QCD), the gauge theory of the strong interaction, is well tested in the high-energy regime where the running coupling is small and perturbative methods are applicable.  At low energies, however, the interaction is strongly coupled and perturbation theory breaks down: quarks and gluons are confined into color-singlet hadrons such as mesons and baryons, and the rich spectrum of resonances arises as excitations of these composite degrees of freedom. Understanding the properties of these resonances, including possible exotic candidates beyond simple $q\bar q$ configurations, is a central goal of hadron spectroscopy. For a recent review see Ref.~\cite{Mai:2022eur}. In this nonperturbative regime, multi-hadron scattering amplitudes provide essential information on color confinement and the internal structure of hadrons.

A prominent example is provided by the COMPASS experiment at CERN, which has performed one of the most detailed amplitude analyses of the $\pi^-\pi^-\pi^+$ system to date~\cite{COMPASS:2015gxz}.  Using high-statistics data from the diffractive reaction $\pi^- p \to \pi^- \pi^- \pi^+ p_{\text{recoil}}$ at $190~\mathrm{GeV}/c$, COMPASS reconstructed the $\pi\pi\pi$ mass spectrum through partial-wave analysis (PWA).  Each partial wave was extracted from the angular distribution of the final-state pions as a function of invariant mass and momentum transfer, revealing detailed information about the underlying $\pi\pi\pi$ interactions and  subsystem $\pi\pi$ interactions. The three-pion final state contains several overlapping resonances and provides access to nontrivial three-body dynamics, including both well-established light-meson states such as the $a_1(1260)$, $a_2(1320)$ and $\pi_2(1670)$, and potential exotic resonances with spin-exotic quantum numbers, for example the $\pi_1(1600)$. In addition to the dominant $a_1(1260)\to\rho\pi$ contribution, the $\pi\pi\pi$ spectrum shows a comparatively narrow enhancement in the $f_0(980)\pi$, $J^{PC}=1^{++}$ channel near $1.42~\text{GeV}$, identified as the $a_1(1420)$~\cite{COMPASS:2020yhb,COMPASS:2015gxz,COMPASS:2018uzl}. In a recent BELLE analysis~\cite{Rabusov:2023tna, Rabusov:2024koz} similar enhancement in tau decays was observed.

The narrow width and the distinct phase motion of the COMPASS results indicate a nontrivial origin within the $\pi\pi\pi$ scattering, and have motivated interpretations in terms of dynamical effects rather than a new genuine resonance state. An example of this is a recent study~\cite{Yan:2025bez} suggesting the origin of the enhancement as a virtual state. Previous studies have interpreted this structure in terms of a triangle singularity (TS) mechanism, in which the $a_1(1260)$ resonance decays into $K^*\bar{K}$, followed by $K^*\to K\pi$ with the pion acting as a spectator and the $K\bar{K}$ pair forming the $f_0(980)$ resonance~\cite{Mikhasenko:2015vca, Mikhasenko:2015oxp, Aceti:2016yeb, Guo:2019twa, Sakthivasan:2024uwd, COMPASS:2020yhb}. Triangle singularities are nonanalytic structures of scattering amplitudes that arise when all internal lines of a triangle diagram can fulfill the on-shell and classical propagation conditions, as first analyzed by Landau~\cite{Landau:1959fi}. This kinematic configuration generates a logarithmic branch point in the complex energy plane close to the physical region, producing a resonance-like enhancement in the production amplitude for the $f_0(980)\pi$ channel without introducing a genuine pole in the scattering amplitude.

Triangle singularities have also been proposed to explain several other narrow or threshold structures in hadron spectroscopy, without any resonance being responsible for them~\cite{Coleman:1965xm, Karplus:1958zz, Landau:1959fi, Booth:1961zz, Anisovich:1964ikk}. See Refs.~\cite{Wu:2011yx,Du:2019idk,Du:2021zdg,Du:2022nno,Cheng:2024sus} for more recent works involving TS. See also Refs.~\cite{Liu:2015taa, Bayar:2016ftu} for the conceptual treatment of TS.  An extensive catalog of such mechanisms across light, strange, charm and bottom sectors is given in Ref.~\cite{Guo:2019twa}, which summarizes many candidate processes in a comprehensive table.  Notable examples include the $\eta(1405/1475)\to3\pi$ decay anomaly, where a $K\bar{K}^*K$ triangle mechanism has been proposed to account for the observed anomalously large isospin violation~\cite{DM2:1989xqc, MARK-III:1990wgk,BES:1998bgh, BES:2000adm, BESIII:2012aa} including the notable application of a unitary three-body formalism~\cite{Nakamura:2023hbt}. 

In Ref.~\cite{Doring:2025phq} a unitary coupled-channel three-body formalism was developed for the strangeness sector. The elusive $K(1460)$ excited kaon~\cite{MartinezTorres:2011gjk} was found to be caused by a triangle singularity at the $3K$ threshold. While near-on-shell kaon exchange is strong enough to generate a resonance pole in the approximation of zero-width $a_0$ and $f_0$ isobars~\cite{Zhang:2021hcl}, the pole disappears for physical $a_0, f_0$ widths onto a hidden Riemann sheet; yet, the interesting question remains in which circumstances triangle singularities can lead to the formation of dynamically generated poles once they are understood as the one-loop approximation of a nonlinear rescattering series. In the charmonium sector, the charged structure $Z_c(3900)$ observed in $e^+e^-\to J/\psi\,\pi^+\pi^-$ can be generated by triangle diagrams with intermediate $D_1\bar{D}D^*$ or $D_0^*(2400)\bar{D}D^*$ mesons going on  shell~\cite{Belle:2013yex, BESIII:2013ris, Szczepaniak:2015eza, Szczepaniak:2015hya, Wang:2013hga, Wang:2013cya}; in the bottomonium sector, the $Z_b(10610)$ structure observed in $e^+e^-\to \Upsilon(1S)\pi^+\pi^-$ can be associated with triangle diagrams involving $B_J^*\bar{B}^*B$ intermediate states~\cite{Belle:2011aa, Szczepaniak:2015eza}; see Refs.~\cite{Dai:2018rra, Liang:2019jtr, Jing:2019cbw, Guo:2019qcn, Molina:2020kyu,Fu:2026ouv} for further examples. These examples illustrate that kinematic singularities can mimic or distort resonance signals in both light and heavy hadron sectors, underscoring the need for frameworks that treat rescattering and channel coupling in a controlled way, particularly for three-body final states.

Most existing studies of triangle singularities are based on explicit triangle loop models and do not implement full three-body unitarity or dynamical coupled-channel rescattering.  Recent theoretical progress has therefore focused on three-body coupled-channel approaches that treat rescattering and channel coupling while maintaining three-body unitarity~\cite{Doring:2025sgb, Feng:2024wyg, Mai:2017bge,Mai:2018djl, Mai:2017vot, Sadasivan:2021emk}. 
Such approaches are also closely related to the development of the three-body quantization conditions needed to extract resonance parameters from the finite-volume spectrum of modern lattice QCD calculations, as demonstrated for the $\pi(1300)$~\cite{Yan:2025mdm}, $\omega(782)$~\cite{Yan:2024gwp,Yu:2026qlt}, and $a_1(1260)$~\cite{Mai:2021nul}. See also Refs.~\cite{Hansen:2019nir,Mai:2021lwb} for dedicated reviews and Ref.~\cite{Sharpe:2026mtt} for a recent status overview of these efforts. In these frameworks, two-body isobars interact with a spectator through iterated rescattering encoded in integral equations that enforce three-body unitarity. The production of an isobar–spectator pair is dressed by all possible rescattering processes, including transitions between different channels and exchanges of quantum numbers among the decay products.
The two-body dynamics enters as input to the three-body equations, for example, through the subsystems $\rho,K^*,\sigma,f_0$,etc. that might be of coupled-channel nature themselves. The isobar dynamics is consistently embedded into a unitary three-body formalism.  As a result, such formalisms naturally generate dynamical thresholds, cusp effects and triangle singularities while preserving the analytic and unitary properties of the full amplitude, providing a framework to study the role of triangle singularities in three-body final states.  

Among the available three-body approaches, we employ the coupled-channel formalism of Ref.~\cite{Feng:2024wyg}, which implements three-body unitarity for the $\pi\pi\pi/\pi K\bar K$ system with nine coupled isobar–spectator channels.  In this work, we revisit the COMPASS $\pi^-\pi^-\pi^+$ data in the region of the $a_1(1420)$ and apply this framework to the $\pi f_0(980)$ channel in $P$-wave, where the $f_0(980)$ decays not only into $K\bar K$ but also into $\pi\pi$ in the final state.  We demonstrate that the triangle enhancement near $1.42~\text{GeV}$ is reproduced, which corresponds to earlier studies~\cite{Mikhasenko:2015vca, Mikhasenko:2015oxp, Aceti:2016yeb, Guo:2019twa, Sakthivasan:2024uwd}, in the coupled channel three-body dynamics without introducing any additional $a_1(1420)$ resonance pole or contact term. A fit to the COMPASS data is performed~\cite{COMPASS:2015gxz, COMPASS:2015kdx, COMPASS:2020yhb} providing strong evidence that the observed structure is kinematic in origin rather than a new resonance state. This highlights the essential role of coupled-channel dynamics in unitary three-body frameworks on the one hand, and the feasibility to embed one-loop TS calculations into unitary amplitudes allowing to consistently incorporate final-state interactions and their potentially substantial effect.

\section{Formalism}
\label{sec:Formalism}
\subsection{The Infinite Volume Unitarity (IVU) framework}
The three-body unitary formalism we refer to as infinite volume unitarity (IVU) was introduced in Ref.~\cite{Mai:2017vot} and later used to study the lineshape and the pole position of the $a_1(1260)$~\cite{Sadasivan:2020syi, Sadasivan:2021emk}. The non-zero strangeness case was explored in Ref.~\cite{Doring:2025phq}. Recently, a nine-channel extension of the method including pions and kaons was discussed in Ref.~\cite{Feng:2024wyg}. This formalism is used in this work to study the $a_1(1420)$ as summarized in the following. In order to describe the interaction of three particles, one classifies the three-particle states through all possible two-particle subsystems --the isobars--, and the third particle --the spectator--, which is on-shell. This follows Faddeev's analysis of three-particle scattering in spirit, see Ref.~\cite{Faddeev:1960su}, and the comparison in Ref.~\cite{Doring:2025phq}. A historical development of similar analyses can also be found in Ref.~\cite{Mai:2017vot}. 
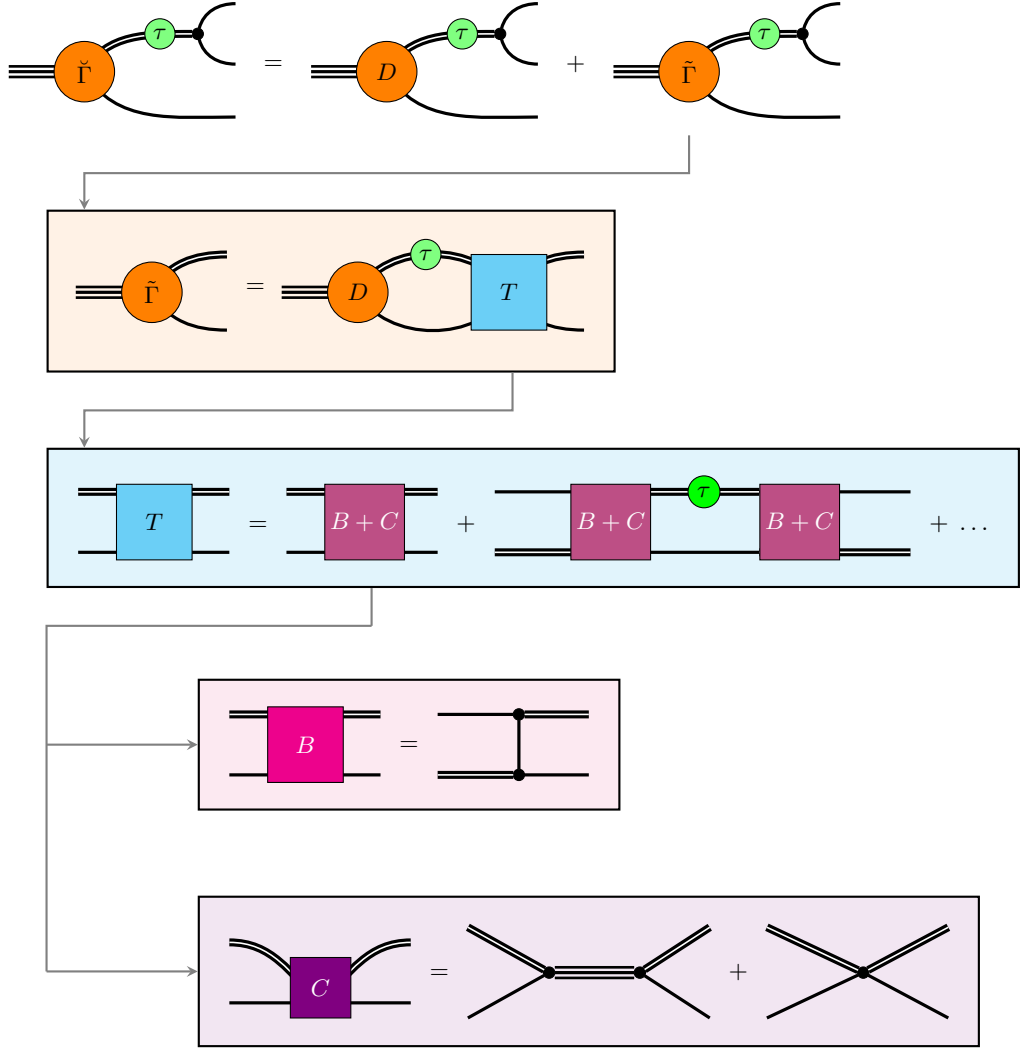
\begin{figure}[t]
    \centering
    \begin{tikzpicture}[x=1cm, y=1cm, >=stealth]
        \node (R1_lhs) at (-3, 0) {
            \begin{tikzpicture}[baseline=(centre.base)]
                \coordinate (centre) at (0,0);
                \begin{feynman}
                    \vertex (piout) at (2,-0.6);
                    \vertex (a) at (0,0);
                    \vertex (Rin) at (-1,0);
                    \vertex[dot] (isoout) at (1.5,0.5) {};
                    \vertex (piouttwo) at (2,0.9);
                    \vertex (pioutthree) at (2,0.1);
                    \diagram*{  
                        (Rin) -- [triple] (a),
                        (a) -- [plain,out=-60,in=180] (piout),
                        (a) -- [double,out=60,in=180] (isoout),
                        (isoout) -- [plain,out=75,in=180] (piouttwo),
                        (isoout) -- [plain,out=-75,in=180] (pioutthree)
                    };
                    \node[draw, fill=green!50, circle, minimum size=4mm, inner sep=0pt] at (1.0,0.5) {\( \tau \)};
                    \node[draw, fill=orange, circle, minimum size=8mm, inner sep=0pt] at (0,0) {\( \breve\Gamma \)};
                 \end{feynman}
            \end{tikzpicture}
        };

        \node (R1_eq) at (-1, 0) {$=$};

        \node (R1_t1) at (1, 0) {
            \begin{tikzpicture}[baseline=(centre.base)]
                \coordinate (centre) at (0,0);
                \begin{feynman}
                    \vertex (Rin) at (-1,0);         
                    \vertex (a) at (0,0);
                    \vertex (piout) at (2,-0.6);
                    \vertex[dot] (isobarout) at (1.5,0.5) {};
                    \vertex (piouttwo) at (2,0.9);
                    \vertex (pioutthree) at (2,0.1);
                    \diagram*{  
                        (Rin) -- [triple] (a),
                        (a) -- [plain,out=-60,in=180] (piout),
                        (a) -- [double,out=60,in=180] (isobarout), 
                        (isobarout) -- [plain,out=75,in=180] (piouttwo),
                        (isobarout) -- [plain,out=-75,in=180] (pioutthree)
                    };
                    \node[draw, fill=green!50, circle, minimum size=4mm, inner sep=0pt] at (1.0,0.5) {\( \tau \)};
                    \node[draw, fill=orange, circle, minimum size=8mm, inner sep=0pt] at (a) {\( D \)};
                \end{feynman}
            \end{tikzpicture}
        };

        \node (R1_plus) at (3, 0) {$+$};

        \node (R1_t2) at (5, 0) {
            \begin{tikzpicture}[baseline=(centre.base)]
                \coordinate (centre) at (0,0);
                \begin{feynman}
                    \vertex (piout) at (2,-0.6);
                    \vertex[dot] (isobarout)  at (1.5,0.5) {};
                    \vertex (a) at (0,0);
                    \vertex (Rin) at (-1,0);
                    \vertex (piouttwo) at (2,0.9);
                    \vertex (pioutthree) at (2,0.1);
                    \diagram*{  
                        (Rin) -- [triple] (a),
                        (a) -- [plain,out=-60,in=180] (piout),
                        (a) -- [double,out=60,in=180] (isobarout),
                        (isobarout) -- [plain,out=75,in=180] (piouttwo),
                        (isobarout) -- [plain,out=-75,in=180] (pioutthree)
                    };
                    \node[draw, fill=green!50, circle, minimum size=4mm, inner sep=0pt] at (1.0,0.5) {\( \tau \)};
                    \node[draw, fill=orange, circle, minimum size=8mm, inner sep=0pt] at (a) {\( \tilde\Gamma \)};
                \end{feynman}
            \end{tikzpicture}
        };

        \node[draw=black, thick, fill=orange!10, inner sep=10pt, anchor=west] (BoxGamma) at (-4, -3) {
            \begin{tikzpicture}[baseline=(centre.base)]
                \coordinate (centre) at (0,0);
                \begin{feynman}
                    \vertex (piout) at (+1,-0.5);
                    \vertex (isobarout)  at (+1,+0.5);
                    \vertex (a) at (0,0);
                    \vertex (Rin) at (-1,0);
                    \diagram*{  
                        (Rin) -- [triple] (a),
                        (a) -- [plain,out=-60,in=180] (piout),
                        (a)-- [double,out=60,in=180] (isobarout), 
                    };
                    \vertex[draw, fill=orange, circle, minimum size=8mm, inner sep=0pt] at (-0.4,0) {\( \tilde\Gamma \)};
                \end{feynman}
            \end{tikzpicture}
            $\;=\;$
            \begin{tikzpicture}[baseline=(centre.base)]
                \coordinate (centre) at (0,0);
                \begin{feynman}
                    \vertex (piout) at (+3,-0.5);
                    \vertex (isobarout)  at (+3,+0.5);
                    \vertex (a) at (0,0);
                    \vertex (b) at (2,0);
                    \vertex (Rin) at (-1,0);
                    \diagram*{  
                        (Rin) -- [triple] (a),
                        (a) -- [plain,out=-60,in=-120] (b),
                        (a)-- [double,out=60,in=120] (b), 
                        (b)-- [double,out=60,in=180] (isobarout),
                        (b) -- [plain,out=-60,in=180] (piout),
                    };
                    \node[draw, fill=cyan!50, rectangle, minimum size=10mm, inner sep=2pt] at (1.5,0) {$T$};
                    \node[draw, fill=green!50, circle, minimum size=4mm, inner sep=0pt] at (0.7,0.5) {\( \tau \)};
                    \node[draw, fill=orange, circle, minimum size=8mm, inner sep=0pt] at (-0.4,0) {\( D \)};
                \end{feynman}
            \end{tikzpicture}
        };

        \node[draw=black, thick, fill=cyan!10, inner sep=10pt, anchor=west] (BoxT) at (-4, -6) {
            \begin{tikzpicture}[baseline=(T.base)]
                \coordinate (centreup) at (0,0.4);
                \coordinate (centredown) at (0,-0.4);
                \begin{feynman}
                    \vertex (isoin) at (-1, 0.4);
                    \vertex (isoout) at (1, 0.4);
                    \vertex (specin) at (-1, -0.4);
                    \vertex (specout) at (1, -0.4);
                    \diagram*{
                        (isoin) -- [double] (centreup) -- [double] (isoout),
                        (specin) -- [plain] (centredown) -- [plain] (specout)
                    };
                    \node[draw, fill=cyan!50, rectangle, minimum size=10mm, inner sep=2pt] (T) at (-0.5,0) {\( T \)};
                \end{feynman}
            \end{tikzpicture}
            $\;=\;$
            \begin{tikzpicture}[baseline=(T.base)]
                \coordinate (centreup) at (0,0.4);
                \coordinate (centredown) at (0,-0.4);
                \begin{feynman}
                    \vertex (isoin) at (-1, 0.4);
                    \vertex (isoout) at (1, 0.4);
                    \vertex (specin) at (-1, -0.4);
                    \vertex (specout) at (1, -0.4);
                    \diagram*{
                        (isoin) -- [double] (centreup) -- [double] (isoout),
                        (specin) -- [plain] (centredown) -- [plain] (specout)
                    };
                    \node[draw, fill=magenta!50!violet, rectangle, text=white, minimum size=10mm, inner sep=2pt] (BC) at (-0.5,0) {\( B+C \)};
                \end{feynman}
            \end{tikzpicture}
            $\;+\;$
            \begin{tikzpicture}[baseline=(T.base)]
                \coordinate (centreup) at (0,0.4);
                \coordinate (centredown) at (0,-0.4);
                \begin{feynman}
                    \vertex (isoin) at (-3, -0.4);
                    \vertex (specin) at (-3, 0.4);
                    \vertex (isoout) at (2.5, -0.4);
                    \vertex (specout) at (2.5, 0.4);
                    \vertex (isointleft) at (-1,0.4);
                    \vertex (isointright) at (0.5,0.4);
                    \vertex (specintleft) at (-1,-0.4);
                    \vertex (specintright) at (0.5,-0.4);
                    \diagram*{
                        (specin) -- [plain] (isointleft) -- [double] (isointright) -- [plain] (specout),
                        (isoin) -- [double] (specintleft) -- [plain] (specintright) -- [double] (isoout)
                    };
                    \node[draw, fill=magenta!50!violet, rectangle, text=white, minimum size=10mm, inner sep=2pt] (BCin) at (-2,0) {\( B+C \)};
                    \node[draw, fill=magenta!50!violet, rectangle, text=white, minimum size=10mm, inner sep=2pt] (BCout) at (0.5,0) {\( B+C \)};
                    \node[draw=black, fill=green, circle, text=black, inner sep=2pt] at (-0.45,0.4) {\( \tau \)};
                \end{feynman}
            \end{tikzpicture}
            $\;+\; \dots$
        };

        \node[draw=black, thick, fill=magenta!10, inner sep=10pt, anchor=west] (BoxB) at (-2, -9) {
            \begin{tikzpicture}[baseline=-0.5ex]
                \coordinate (centreup) at (0,0.4);
                \coordinate (centredown) at (0,-0.4);
                \begin{feynman}
                    \vertex (isoin) at (-1, 0.4);
                    \vertex (isoout) at (1, 0.4);
                    \vertex (specin) at (-1, -0.4);
                    \vertex (specout) at (1, -0.4);
                    \diagram*{
                        (isoin) -- [double] (centreup) -- [double] (isoout),
                        (specin) -- [plain] (centredown) -- [plain] (specout)
                    };
                    \node[draw, fill=magenta, rectangle, text=white, minimum size=10mm, inner sep=2pt] (B) at (-0.5,0) {\( B \)};
                \end{feynman}
            \end{tikzpicture}
            $\;=\;$
            \begin{tikzpicture}[baseline=-0.5ex]
                \begin{feynman}
                    \vertex (in_t) at (-1, 0.4);
                    \vertex (in_b) at (-1, -0.4);
                    \vertex[dot] (t) at (0, 0.4) {};
                    \vertex[dot] (b) at (0, -0.4) {};
                    \vertex (out_t) at (1, 0.4);
                    \vertex (out_b) at (1, -0.4);
                    \diagram*{
                        (in_t) -- [plain] (t) -- [double] (out_t),
                        (in_b) -- [double] (b) -- [plain] (out_b),
                        (t) -- [plain] (b)
                    };
                \end{feynman}
            \end{tikzpicture}
        };

        \node[draw=black, thick, fill=violet!10, inner sep=10pt, anchor=west] (BoxC) at (-2, -12) {
            \begin{tikzpicture}[baseline=-0.5ex]
                \begin{feynman}
                    \vertex (in_t) at (-1.2, 0.4);
                    \vertex (in_b) at (-1.2, -0.4);
                    \vertex (out_t) at (1.2, 0.4);
                    \vertex (out_b) at (1.2, -0.4);
                    \vertex (c) at (0,-0.4);
                    \diagram*{
                        (in_t) -- [double, out=0, in=135] (c),
                        (in_b) -- [plain] (c),
                        (c) -- [double, out=45, in=180] (out_t),
                        (c) -- [plain] (out_b)
                    };
                    \node[draw=black, fill=violet, rectangle, text=white, inner sep=4pt, minimum size=8mm] at (-0.4,-0.2) {\( C \)};
                \end{feynman}
            \end{tikzpicture}
            $\;=\;$
            \begin{tikzpicture}[baseline=-0.5ex]
                \begin{feynman}
                    \vertex (in_t) at (-1.6, 0.6);
                    \vertex (in_b) at (-1.6, -0.6);
                    \vertex[dot] (v1) at (-0.6, 0) {};
                    \vertex[dot] (v2) at (0.6, 0) {};
                    \vertex (out_t) at (1.6, 0.6);
                    \vertex (out_b) at (1.6, -0.6);
                    \diagram*{
                        (in_t) -- [double] (v1),
                        (in_b) -- [plain] (v1),
                        (v1) -- [triple] (v2),
                        (v2) -- [double] (out_t),
                        (v2) -- [plain] (out_b)
                    };
                \end{feynman}
            \end{tikzpicture}
            $\;+\;$
            \begin{tikzpicture}[baseline=-0.5ex]
                \begin{feynman}
                    \vertex (in_t) at (-1.2, 0.6);
                    \vertex (in_b) at (-1.2, -0.6);
                    \vertex[dot] (v) at (0, 0) {};
                    \vertex (out_t) at (1.2, 0.6);
                    \vertex (out_b) at (1.2, -0.6);
                    \diagram*{
                        (in_t) -- [double] (v),
                        (in_b) -- [plain] (v),
                        (v) -- [double] (out_t),
                        (v) -- [plain] (out_b)
                    };
                \end{feynman}
            \end{tikzpicture}
        };

        \draw[gray, thick, ->] ([xshift=-0.5cm]R1_t2.south) -- ++(0,-0.5) -| ([xshift=0.5cm]BoxGamma.north west);

        \draw[gray, thick, ->] ([xshift=2.4cm]BoxGamma.south) -- ++(0,-0.5) -| ([xshift=0.5cm]BoxT.north west);

        \draw[gray, thick] ([xshift=4.3cm]BoxT.south west) -- ++(0,-0.5) coordinate (spineT);
        \draw[gray, thick] (spineT) -| (-4, -12);
        \draw[gray, thick, ->] (-4, -9) -- (BoxB.west);
        \draw[gray, thick, ->] (-4, -12) -- (BoxC.west);
    \end{tikzpicture}%
    \caption{The schematics of the infinite volume unitarity formalism. The first row depicts the quantity $\breve\Gamma$, which corresponds to the amplitude decomposed into disconnected and connected components. The second row shows the connected component of the amplitude given by the quantity $\tilde\Gamma$. In the third row, the diagrammatic representation of the Lippmann-Schwinger-like equation is shown, which is parametrized by the $B$-, $C$- and the $\tau$-terms. The last two rows show the $B$-term, which describes the long-range interaction between a spectator and an isobar constrained by unitarity, as well as the $C$-term, which describes the contact interactions and possible simple poles not constrained by unitarity. The simple pole is denoted by the triple line here.}
    \label{fig:IVU-schematic}
\end{figure}
A schematic representation of the formalism is given in Fig.~\ref{fig:IVU-schematic}. The isobars and the spectators interact through exchange of a particle and this is given by the $B$-term denoted by the magenta box in the second-to-last row in Fig.~\ref{fig:IVU-schematic}. The two-particle dynamics are described the $\tau$-term denoted by the green circles. Note that throughout this paper we omit the ``tilde'' symbol used for $B$ and $\tau$ in previous papers~\cite{Feng:2026ixm, Doring:2025phq, Feng:2024wyg}, simply to de-clutter the notation. The distribution of symmetry and isospin factors among $B$ and $\tau$ is the same as in the $\tilde B$ and $\tilde \tau$ of these publications, but different from publications before that, e.g., in Ref.~\cite{Sadasivan:2021emk}.  The isobars have specific quantum numbers, so does a given channel of isobar and spectator. The formalism casts the full interaction in the form of a Lippmann-Schwinger-like equation, which is denoted by the third row in Fig.~\ref{fig:IVU-schematic}. This ensures that the interaction is properly unitarized and the final state interactions are taken into account, which is crucial for a sound theory that describes low-energy QCD. Introducing the labels $i$, $j$ and $k$ to denote the channels, the Lippmann-Schwinger-like equation for the channel transition $i\to j$ is given by
\begin{equation}
    T_{ji}(s,\bm{p'},\bm{p}) = B_{ji}(s,\bm{p'},\bm{p}) + C_{ji}(s,\bm{p'},\bm{p}) + \int\limits  \frac{\text{d}^3 l}{(2\pi)^3\,2E_{l,k}} \left( B_{jk}(s,\bm{p'},\bm{l}) + C_{jk}(s,\bm{p'},\bm{l}) \right)\, \tau_{kk'}(\sigma_{l,k}) \, T_{k'i}(s,\bm{l},\bm{p})\,.
    \label{eq:LS-HB}
\end{equation}
The relevant kinematic variables are: the three-body invariant mass given by the Mandelstam-$s$, the three-momentum of the incoming and outgoing spectators given by $\bm{p}$ and $\bm{p^\prime}$, respectively, the two-body invariant mass squared of the isobar given by $\sigma_{l,k} = s + m_k^2 - 2\sqrt{s}E_{l,k}$, where $E_{l,k} = \sqrt{m_k^2 + \bm{l}^2}$ with $m_k$ denoting the mass of the spectator in channel $k$. Anticipating the channel space given in~\cref{tab:TS channels}, we note here that while $\tau$ can have off-diagonal elements (cf. channel $5$ and $6$), referred to later as an in-flight transition, the spectator remains unchanged. First, the phase space integral in momentum $l$ over the on-shell spectator is carried out up to a cutoff $\Lambda$. Second, there is an additional term $C$ in Eq.~\eqref{eq:LS-HB}. As noted before, unitarity constrains only the long-range (exchange) force between the isobars and the spectators. The additional degrees of freedom are short-range forces parametrized by the $C$-term, see also Fig.~\ref{fig:IVU-schematic}. Besides the usual contact terms that are analytic in the intrinsic invariant mass variable of the system, the $C$-term can also include simple poles, which can be fit to study resonance poles, see, e.g., Refs.~\cite{Sadasivan:2020syi, Sadasivan:2021emk}. The quantities discussed so far describe only the interactions between the isobars and the spectators. In order to translate this to study a process which describes a source producing an isobar and a spectator, we include a term which describes the dissociation of the source into an isobar and a spectator given by the $D$-term, which is symbolized in Fig.~\ref{fig:IVU-schematic}. This term has to be parametrized and fit to experimental data, as well. The final amplitude after all the rescatterings are taken into account is given by $\tilde\Gamma$, also shown in Fig.~\ref{fig:IVU-schematic},
\begin{equation}
    \tilde\Gamma_j (s,\bm{p}') =\int_{\Gamma}\frac{\diff p\,\bm{p}^2}{(2\pi)^3\,2E_{p,i}}\, T_{ji}(s,\bm{p}',\bm{p})\tau_{ik}(\sigma(p))D_k(s,\bm{p})\,,
    \label{eq: gammatilde (TS) old}
\end{equation}
Unitarity of the full production amplitude $\breve\Gamma$ demands that the fully disconnected term in the first row in Fig.~\ref{fig:IVU-schematic} must also be included. Therefore, the full production amplitude is given by $\breve\Gamma$,
\begin{equation}
    \breve\Gamma_j(s,\bm{p'}) = v_j\ \tau_{jk}(\sigma(p^\prime))[D_k(s,\bm{p}') + \tilde\Gamma_k(s, \bm{p}')]\,.
    \label{eq:gammabreve}
\end{equation}
Note the presence of the final decay vertex $v$ in $\breve\Gamma$ (see also \cref{fig:IVU-schematic}). This introduces an additional angle dependence into the amplitude, with the overall angular structure discussed, e.g., in Refs.~\cite{JPAC:2018zwp, Sadasivan:2020syi, Sadasivan:2021emk}. In the present paper we only compare to partial waves extracted from experiment, for a channel with scalar isobar decays, for which $v=1$ as explained in the next subsection, i.e., in practical terms there is no difference if the final decay vertex is included or not. In summary, three-body unitarity precisely fixes Im $B^{-1}$~\cite{Mai:2017vot}, and the $\tau$-term is determined purely by two-body dynamics and  constrained by two-body unitarity. The quantities $C$ and $D$ are fit to experimental data.

\subsection{Channel Space and Transitions}
The exchange interaction between a spectator and an isobar is described by the $B$-term. The spectators considered here are pions and kaons, which are spinless.
However, the isobars can have an overall spin. We define the $B$-term in the helicity basis (HB) first, which makes the handling of various channels manageable. For an incoming and an outgoing on-shell spectator with momentum $p$ and $p^\prime$, respectively, and a total four-momentum $P_3=(\sqrt{s},\bm{0})$, the $B$-term is given by
\begin{align} 
    B_{ji}(s,p^\prime,p)&
    =\frac{({\bm I}_F)_{ji}\,
    v_j^*(p,P_3-p-p')
    v_i(p',P_3-p-p')}
    {2E_{p'+p}(\sqrt{s}-E_{p}-E_{p'}-E_{p'+p})+i\epsilon}
    \quad \text{for} \quad
    i,j\in {\rm HB}\,,
    \label{eq:B-term}
\end{align}
with the angular structures of the isobars with a non-zero total spin contained in the vertices, which are defined as
\begin{align}
    v_{i}(p,q)
    =
    \left\{
   \begin{matrix}
    \begin{array}{ll}
        (p-q)^{\mu}\,\epsilon_{\mu}(\bm{p}+\bm{q}, \lambda)
        \hfill &\quad \text{for $\ell=1$ isobars\,,}\\
        1&\quad \text{for $\ell=0$ isobars\,.}
    \end{array}
    \end{matrix}
    \right.
    \label{eq:v feng2024}
\end{align}
Note that this already describes a transition from channel $i$ to channel $j$. The polarization vectors that enter the vertices for the isobars with a non-zero spin are given in App.~\ref{app:pol}. Further, the factor $\bm{I}_F$ are the isospin coefficients that were derived in Ref.~\cite{Feng:2024wyg}. In this work, we consider the $11$ channels in helicity basis given in the second row of Tab.~\ref{tab:TS channels}. Note that certain channels are ignored due to negligible inelasticities. The projection of the $B$-term on to the $JLS$ basis is given by
\begin{equation}
    B_{ji}(s,p^\prime,p) = 2\pi \int_{-1}^1 dz\ d^J_{\lambda(j)\lambda(i)}(z) B_{ji}(s,\bm{p^\prime},\bm{p})\,,
\end{equation}
where $z=\cos\theta$ is the cosine of the scattering angle between $\bm{p^\prime}$ and $\bm{p}$, and $\lambda(i)$, $\lambda(j)$ are the helicities of the isobars in channels $i$ and $j$, respectively. The $d$ in the above equation is the Wigner-$d$, which is related to the Wigner-$D$ through $d^J_{\lambda\lambda^\prime}(\theta) = D^J_{\lambda\lambda^\prime}(0,\theta,0)$. 
As explained in Ref.~\cite{Feng:2024wyg} the above convention of having the outgoing helicity in the first argument of the Wigner function implies that vector mesons entering in the $z$-direction. The $B$-term for channels with specific angular momentum is derived by including the relevant Clebsch-Gordan coefficients, and is given by
\begin{equation}
    B_{L^\prime(j) L(i)}(s,p',p) = U_{L^\prime(j) \lambda(j)} B_{ji}(s,p',p)U_{\lambda(i)L(i)}\quad \text{for} \quad L,L'\in JLS\,,
    \label{eq:helicity to JLS}
\end{equation}
where
\begin{equation}
    U_{L\lambda }
    =\sqrt{\frac{2L+1}{2J+1}} \sum_{\lambda_2}\cg{L0}{S\lambda}{J\lambda}\cg{S_1\lambda_1}{S_2-\lambda_2}{S\lambda} 
    = \sqrt{\frac{2L+1}{2J+1}}\cg{L0}{S\lambda}{J\lambda}\,.
    \label{eq: HB to JLS}
\end{equation}
Here, $S_1=S$ ($\lambda_1=\lambda$) is the spin (helicity) of the isobar and $S_2=0$ ($\lambda_2=0$) is the spin (helicity) of the spectator. Note that for the channels $\pi\rho$ and $KK^*$ with  $P$-wave isobars,
\begin{align}
    U=
    \begin{pmatrix}
        \frac{1}{\sqrt{3}}  & \frac{1}{\sqrt{3}}  &\frac{1}{\sqrt{3}}  \\
         \frac{1}{\sqrt{6}}  &-\sqrt{\frac{2}{3}}   &\frac{1}{\sqrt{6}} 
    \end{pmatrix} \,,
    \label{Umatrix}
\end{align}
where the first and second rows denote $S$- and  $D$-wave, respectively. For channels without spin this reduces to $U=1$. The $9$ channels in the $JLS$ basis considered in this work are given in the third row of Tab.~\ref{tab:TS channels}. The isospin factors for the channel transitions are given in Tab.~\ref{tab:isospin}, which is symmetric. In the $JLS$ basis, the $\pi\rho$ and the $K K^*$ channels couple to $J=1$ in relative $S$- and $D$-waves, whereas all the other channels, which are spinless, are in relative $P$-waves. Note that the strange and the non-strange sectors are disconnected, and a transition across these sectors can only happen through an in-flight transition of the isobar in channels $5$ and $6$. This is discussed in the next section. In $JLS$ basis, the Lippmann-Schwinger-like equation reads
\begin{equation}
    T_{ji}(s,p^\prime,p) = B_{ji}(s,p^\prime,p) + C_{ji}(s,p^\prime,p) + \int\limits_0^\Lambda \frac{\text{d}l\,l^2}{(2\pi)^3\,2E_{l,k}} \left( B_{jk}(s,p^\prime,l) + C_{jk}(s,p^\prime,l) \right)\, \tau_{kk'}(\sigma_l) \, T_{k'i}(s,l,p)\,,
    \label{eq:LS-JLS}
\end{equation}
where the indices now denote the channels in the $JLS$ basis which allows one to drop the explicit $L$ and $L^\prime$ indices, see Tab.~\ref{tab:TS channels}. Correspondingly, the 9-channel, partial-wave projected production amplitude is obtained from \cref{eq: gammatilde (TS) old} and reads
\begin{align}
        \tilde\Gamma_j (s,p') =\int_{\Gamma}\frac{\diff p\,p^2}{(2\pi)^3\,2E_{p,i}}\, T_{ji}(s,p',p)\tau_{ik}(\sigma(p))D_k(s,p)\,.
    \label{eq:gammatilde_PW}
\end{align}
\begin{table}[t]
    \centering
    \caption{ 
    Channel numbers and associated $JLS$, helicity basis (HB) and quantum numbers of the involved isobars.
    The abbreviation $\pi_2$ denotes the repulsive $S$-wave $\pi\pi$ channel with isobar isospin $I_I=2$, and $K_{3/2}$ denotes the repulsive $S$-wave $\pi K$ channel with isobar isospin $I_I=\frac{3}{2}$. In the $(\ell,I_I)=(0,0)$ sector, the isobar has two modes, $\pi\pi$ and $K\bar K$, which are counted as separate three-body channels.
    }
    \label{tab:TS channels}
    \setlength{\tabcolsep}{7pt}
    \begin{adjustbox}{max width=\linewidth}
    \renewcommand{\arraystretch}{1.2}
        \begin{tabular}{l|c|c|c|c|c|c|c|c|c}
            \hline\hline
            Channel number & 1 & 2 & 3 & 4 & 5 & 6 & 7 & 8 & 9 \\
            \hline
            $JLS$ basis
            &
            $(\pi\rho)_S$
            &
            $(\pi\rho)_D$
            &
            $(K K^*)_S$
            &
            $(K K^*)_D$
            &
            $(\pi f_0(500))_P$
            &
            $(\pi f_0(980))_P$
            &
            $(\pi\pi_2)_P$
            &
            $(K K^*_0(700))_P$
            &
            $(K K_{\nicefrac{3}{2}})_P$
            \\ 
            \hline
            Helicity basis
            &
            \multicolumn{2}{c|}{$\pi\rho_{\lambda=\pm 1,0}$}
            &
            \multicolumn{2}{c|}{$K K^*_{\lambda=\pm 1,0}$}
            &
            $\pi f_0(500)$
            &
            $\pi f_0(980)$
            &
            $\pi\pi_2$
            &
            $K \kappa$
            &
            $K K_{\nicefrac{3}{2}}$
            \\
            \hline
            Isobar $(\ell,I_I)$
            &
            \multicolumn{2}{c|}{$(1,1)$}
            &
            \multicolumn{2}{c|}{$(1,\frac{1}{2})$}
            &
            \multicolumn{2}{c|}{$(0,0)$}
            &
            \multicolumn{1}{c|}{$(0,2)$}
            &
            \multicolumn{1}{c|}{$(0,\frac{1}{2})$}
            &
            \multicolumn{1}{c}{$(0,\frac{3}{2})$}
            \\
            \hline \hline
        \end{tabular}
    \end{adjustbox}
\end{table}

\subsection{Isobar-Spectator Propagation and Other Ingredients}
In our formalism, the isobar-spectator propagator $\tau$-term when taken together with the adjacent decay vertices should correspond to the full plane-wave isospin projected $2\to2$ scattering amplitudes without any symmetry factors absorbed in the definition two-body states as it is customary for many amplitude formulations. This exact matching is described in Sec.~\ref{sec:2b input}. And for the channels given in Tab.~\ref{tab:TS channels}, the $\tau$-term is given by the following block diagonal matrix in the $JLS$ basis:
\begin{equation}
    \textrm{diag}
    \begin{pmatrix}
        \tau_{\pi\rho} & \tau_{\pi\rho} & \tau_{KK^*} & \tau_{KK^*} &
        \begin{pmatrix}
            \tau_{\pi\pi \to \pi\pi}^{I=0,J=0} &  \tau_{\pi\pi \to K\bar{K}}^{I=0,J=0} \\
            \tau_{\pi\pi \to K\bar{K}}^{I=0,J=0} & \tau_{K\bar{K} \to K\bar{K}}^{I=0,J=0}
        \end{pmatrix} &
        \tau_{\pi\pi_2} & \tau_{K K^*_0(700)} & \tau_{KK_{\nicefrac{3}{2}}}
    \end{pmatrix}\,.
    \label{eq:tau}
\end{equation}
A given $\tau$ with an isobar containing a resonance corresponds to the isobar matched to the relevant two-body scattering amplitude, e.g., $\tau_{\pi\rho}$ requires one to consider the $I=1$, $L=1$ two-body scattering amplitude. For the cases when the label doesn't include a resonance, $\tau_{\pi\pi_2}$ corresponds to the $I=2$, $L=0$ two-body scattering amplitude and $\tau_{KK_{\nicefrac{3}{2}}}$ corresponds to the $I=3/2$, $L=0$ two-body scattering amplitude. The only exception is the $2\times 2$ matrix in the equation above. These channels allow for an in-flight transition across the two channels. The relevant quantum numbers and the particles are given as labels in this case. 
The $\tau$-term is not affected by the isobar-spectator partial-wave projection discussed before.
\begin{table}[t]
    \centering
    \caption{The isospin factors $\bm{I_F}$ in Eq.~\eqref{eq:B-term}. A given matrix element with a row index $i$ and a column index $j$ denotes a transition from channel $j$ to channel $i$ in the $JLS$ basis, see Tab.~\ref{tab:TS channels}.}
    \label{tab:isospin}
    \setlength{\tabcolsep}{7pt}   
    \begin{adjustbox}{max width=\linewidth}
        \begin{tabular}{c |c c c c c c c c c}
        & $(\pi\rho)_S$ & $(\pi\rho)_D$ & $(K K^*)_S$ & $(K K^*)_D$ & $(\pi f_0(500))_P$ & $(\pi f_0(980))_P$ & $(\pi \pi_2)_P$ & $(K K^*_0(700))_P$ & $(K K_{\nicefrac{3}{2}})_P$ \\ \hline
        $(\pi\rho)_S$ & $\frac{1}{2}$ & $\frac{1}{2}$ & $0$ & $0$ & $\frac{1}{\sqrt{3}}$ & $0$ & $-\sqrt{\frac{5}{12}}$ & $0$ & $0$ \\
        $(\pi\rho)_D$ & & $\frac{1}{2}$ & $0$ & $0$ & $\frac{1}{\sqrt{3}}$ & $0$ & $-\sqrt{\frac{5}{12}}$ & $0$ & $0$ \\
        $(K K^*)_S$ & & & $-\frac{1}{3}$ & $-\frac{1}{3}$ & $0$ & $-\sqrt{\frac{2}{3}}$ & $0$ & $-\frac{1}{3}$ & $-\frac{\sqrt{8}}{3}$ \\
        $(K K^*)_D$ & & & & $-\frac{1}{3}$ & $0$ & $-\sqrt{\frac{2}{3}}$ & $0$ & $-\frac{1}{3}$ & $-\frac{\sqrt{8}}{3}$ \\
        $(\pi f_0(500))_P$ & & & & & $\frac{1}{3}$ & $0$ & $\frac{\sqrt{5}}{3}$ & $0$ & $0$ \\
        $(\pi f_0(980))_P$ & & & & & & $0$ & $0$ & $-\sqrt{\frac{2}{3}}$ & $-\frac{2}{\sqrt{3}}$\\
        $(\pi \pi_2)_P$ & & & & & & & $\frac{1}{6}$ & $0$ & $0$ \\
        $(K K^*_0(700))_P$ & & & & & & & & $-\frac{1}{3}$ & $-\frac{\sqrt{8}}{3}$ \\
        $(K K_{\nicefrac{3}{2}})_P$ & & & & & & & & & $\frac{1}{3}$
        \end{tabular}
    \end{adjustbox}
\end{table}

The only ingredient that has not been discussed up until this point is the contact and the three-body dissociation terms. In this work, the $C$-term is parametrized as
\begin{equation}
    C_{ji}(p,q) = \left(\frac{g_j g_i}{s-m^2}\right) \mathring{D}_{L_j}(q; \lambda_{BW}) \mathring{D}_{L_i}(p; \lambda_{BW}) {H(q; \lambda_H)} {H(p; \lambda_H)}\,.
    \label{eq:C}
\end{equation}
The $C$-term consists only of a simple pole, and the couplings $g$ and the mass $m$ are fitted to the available data. If a pole is not explicitly included in the fits, there is no $C$-term, i.e., non-singular short-range interactions are omitted in the present approach, simply because good fits can be obtained without them. In addition, the $C$-term includes  the Blatt-Weisskopf centrifugal barrier factors given by~\cite{Mai:2021vsw}
\begin{equation}
    \mathring{D}_{L}(p;\lambda_{BW}) =
    \begin{cases}
        1,\quad &L=0,\\
        \frac{p}{\sqrt{\lambda_{BW}^2 p^2 + 1}},\quad &L=1,\\
        \frac{p^2}{\sqrt{\lambda_{BW}^4 p^4 + 3 \lambda_{BW}^2 p^2 + 9}},\quad &L=2\,,
    \end{cases}
    \label{eq:BWfactor}
\end{equation}
where $L$ denotes the relative angular momentum between the isobar and the spectator, $p$ denotes the spectator three-momentum and $\lambda_{BW}$ is the range value. The form factor reads
\begin{equation}
    H(p;\lambda_H) = \frac{\lambda_H^4}{p^4 + \lambda_H^4},
    \label{eq:tamingfactor}
\end{equation}
which ensures the correct large energy behavior. The term $\lambda_H$ is the form factor cutoff. The three-body dissociation vertex $D$-term (see \cref{fig:IVU-schematic}) is parametrized as 
\begin{equation}
    D_i(p) = \left(d_i+d\frac{g_i}{s-m^2}\right) \mathring{D}_{L_i}(p;\lambda_{BW}){H(p; \lambda_H)}\,.
    \label{eq:D}
\end{equation}
The additional factors are the same as before. The term $d_i$ is channel dependent, whereas the term $d$ is channel-independent needed for the unknown, overall normalization. These two terms are also fit to the available data. This concludes all the ingredients that go into the formalism. The exact procedure of solving the Lippmann-Schwinger-like is discussed in Sec.~\ref{sec:solveIVU}.

\section{Implementation}
\subsection{Two-body Input}
\label{sec:2b input}
The two-body dynamics of isobars in the infinite volume unitarity (IVU) formalism is constrained by coupled-channel unitarity and the corresponding right-hand cuts while left-hand cuts, that could generate spurious singularities in the three-body integration, are usually neglected. We need an input for the isobars in the $9$ different channels in the $JLS$ basis given in Tab.~\ref{tab:TS channels}. In a previous work~\cite{Sakthivasan:2024uwd} where the IVU formalism was discussed in the context of Landau singularities, resummed propagators for the two-body amplitudes were matched to Breit-Wigner parameterizations with PDG values for the masses and the widths of the resonances that these channels contain. These propagators were used as input for the isobars. This has the disadvantage that the inelasticity in the channels $5$ and $6$ which contain the $f_0(500)$ and $f_0(980)$ are not taken into account, and the two-body sub-amplitudes do not describe the experimental phase shifts quantitatively. In a more recent work on the IVU formalism~\cite{Feng:2024wyg}, the two-body amplitudes given in Ref.~\cite{Oller:1998hw} were used for scalar isobars corresponding to the channels $5$--$9$, namely the channels containing the $f_0(500)$, $f_0(980)$ and $K^*(700)$. Unlike before, with this input the inelasticities in the channels are properly taken into account. However, because of the high elasticity of the vector channels, resummed propagators were used for channels $1$--$4$, namely the channels containing the $\rho(770)$ and $K^*(892)$, which were fit to experimental data. In this work, for the sake of consistency, we use the two-body amplitudes given in Ref.~\cite{Oller:1998hw} for all the isobars. This is an earlier implementation of the inverse amplitude method (IAM) introduced in Refs.~\cite{Oller:1997ti, Dobado:1996ps} differing from the currently more commonly used IAM~\cite{Guerrero:1998ei, GomezNicola:2001as} in that only the tree-level amplitudes are included at $\mathcal{O}(p^4)$ in chiral perturbation theory (CHPT) and the loop contributions are excluded. We prefer this due to its computational simplicity, and we refer to this as the IAM-OOP or simply the OOP method after its authors.
\begin{figure}[t]
    \centering
    \includegraphics[width=0.95\linewidth]{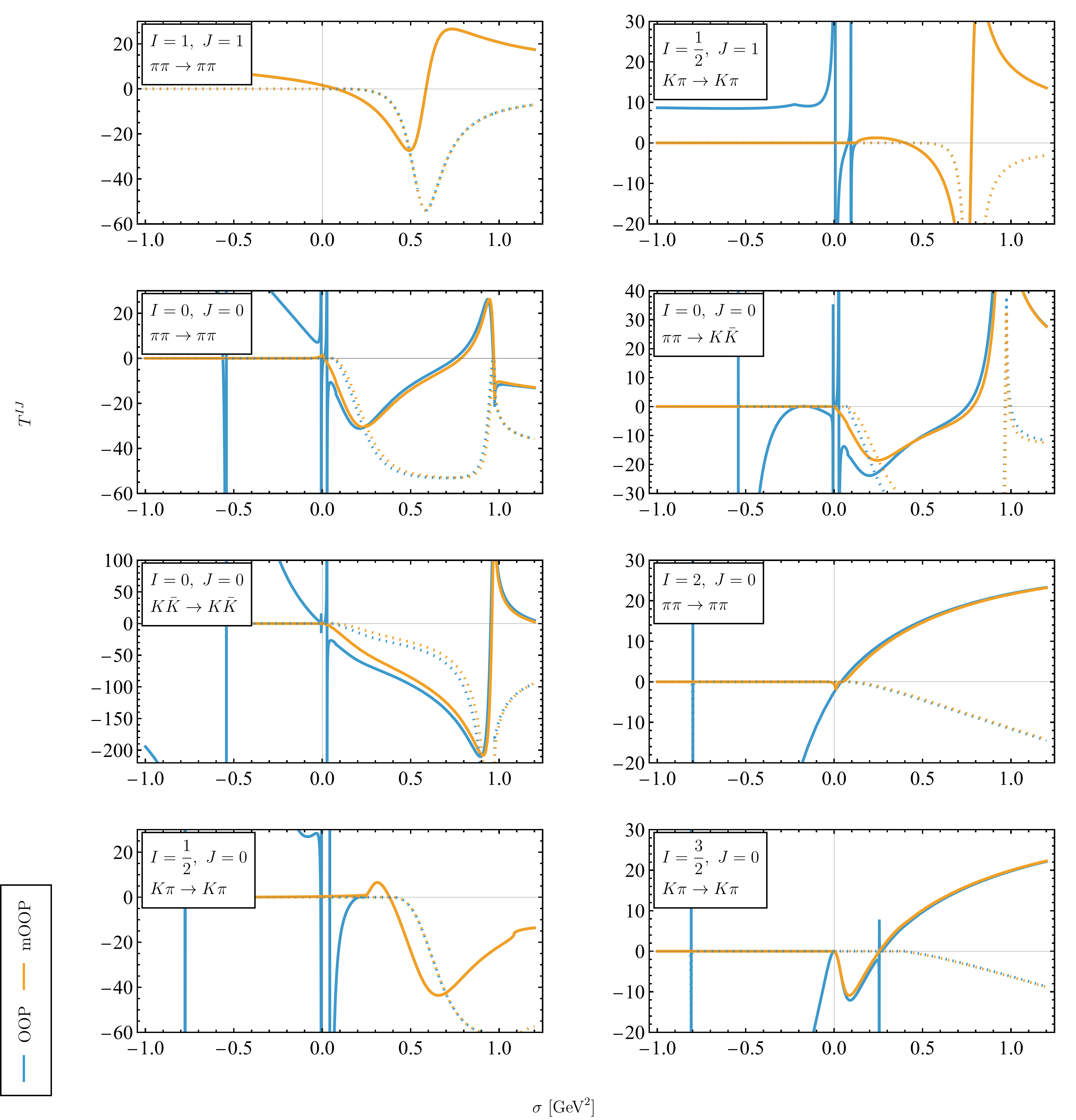}
    \caption{The two-body amplitudes used as an input for the isobars plotted for relevant two-body invariant mass intervals. The OOP method suffers from the existence of spurious poles below threshold. This is remedied by the mOOP method, which is used in this work. The real parts of the amplitudes are given by the solid lines and the imaginary parts are given by the dashed lines.}
    \label{fig:OOP}
\end{figure}%

In this method, the unitarized two-body amplitude for a given channel with total isospin $I$ and total angular momentum $J$ is given by
\begin{equation}
    T^{IJ} = T_2^{IJ} (T_2^{IJ} - T_{4P}^{IJ} - T_2^{IJ} G T_2^{IJ})^{-1} T_2^{IJ}\,,
\end{equation}
where $T_2$ corresponds to the $\mathcal{O}(p^2)$ CHPT contributions to the amplitude and $T_{4P}$ corresponds to the tree-level $\mathcal{O}(p^4)$ CHPT contributions, and $G$ is an analytic two-meson loop function. The amplitudes are always $2\times 2$ matrices owing to their coupled-channel nature, except for the isoscalar $K\pi$ two-body amplitude with $J=3/2$ and the isoscalar $\pi\pi$ two-body amplitude with $J=2$. The relevant isospin amplitudes $T^{I}$ can be found in the original reference, and the amplitudes with definite total angular momentum are given by
\begin{equation}
    T^{IJ} = \frac{1}{2} \int_{-1}^1 d(\cos\theta)\ P_J(\cos\theta) T^{I}(\theta)\,,
\end{equation}
where $P_J$ is the Legendre polynomial. We omit the two-body invariant mass dependence of the amplitudes in the above equations for brevity. These amplitudes are explicitly matched to experimental data through the $S^{IJ}$-matrix given by
\begin{equation}
    S^{IJ} = 
    \begin{pmatrix}
        \eta e^{2i\delta_1} & i\sqrt{1-\eta^2} e^{i(\delta_1 + \delta_2)} \\
        i\sqrt{1-\eta^2} e^{i(\delta_1 + \delta_2)} & \eta e^{2i\delta_2}
    \end{pmatrix}\,.
\end{equation}
The elements of the above matrix given by the subscript $m$, $n$ are related to the coupled channel amplitudes through
\begin{equation}
    \begin{aligned}
        [T]_{11} = -\frac{8\pi\sqrt{\sigma}}{2i p_1} \left( [S]_{11} - 1 \right),&&[T]_{22} = -\frac{8\pi\sqrt{\sigma}}{2i p_2} \left( [S]_{22} - 1 \right),&&[T]_{12} = [T]_{21} = -\frac{8\pi\sqrt{\sigma}}{2i\sqrt{p_1 p_2}} [S]_{12}\,,
    \end{aligned}
\end{equation}
where $p_i$ denotes the magnitude of the center-of-mass (c.m.) three-momentum in the relevant channel. Furthermore, unitarity demands
\begin{equation}
    [\textrm{Im}~(T^{IJ})^{-1}]_{ii} = \frac{p_i}{8\pi\sqrt{s}} \Theta(s - (m_{1,i} + m_{2,i})^2)\,,
\end{equation}
where $m_{1,i}$ and $m_{2,i}$ denote the two-mesons in a given channel. Within the IVU formalism, the isobar component $\tau$ when taken together with the adjacent angle-dependent decay vertices in the exchange term $B$ should correspond to the full definite isospin and angular momentum two-meson scattering amplitude from any consistent method, which is OOP in this case. Note that the amplitudes in OOP also include symmetry factors, which need to be removed from the full amplitude. With that, the matching of the isobar components in the $9$ channels to the OOP two-body amplitudes is given by
\begin{equation}
    \begin{aligned}
        &\tau_{11}=\tau_{22} = \frac{3}{2 p^2} T^{11}(\pi\pi \to \pi\pi),&& \tau_{33}=\tau_{44} = \frac{3}{4 p^2} T^{\frac{1}{2}1}(K\pi \to K\pi)\,,&\\
        &\tau_{55} = 2 T^{00}(\pi\pi \to \pi\pi),&& \tau_{56} = \tau_{65}= \sqrt{2} T^{00}(\pi\pi \to K\bar{K})\,,&\\
        &\tau_{66} = T^{00}(K\bar{K} \to K\bar{K}),&& \tau_{77} = 2T^{20}(\pi\pi\to\pi\pi)\,,\\
        &\tau_{88} = T^{\frac{1}{2}0}(K\pi \to K\pi), && \tau_{99} = T^{\frac{3}{2}0}(K\pi \to K\pi)\,,
    \end{aligned}
    \end{equation}
where the subscripts of $\tau$ denote the channel, see Tab.~\ref{tab:TS channels} and Eq.~\eqref{eq:tau}, and $\tau_{ji}=0$ for all other matrix elements. Note that these relations depend on the definition of isobar vertices of Eq.~\ref{eq:v feng2024}. We follow here the same separation of angular structures from vector isobars into the $B$-term as in Ref.~\cite{Feng:2026ixm}. Again, we can see that only the channels $5$ and $6$ containing the $f_0(500)$ and $f_0(980)$ allow for an in-flight transition between the isobars, requiring the cross-channel OOP transition amplitudes. The $\rho$, $K^*$ and $\kappa$ isobars have very small inelasticities, and we only take the respective diagonal channels $\pi\pi\to\pi\pi$ and  $\pi K\to\pi K$ into account. This results in the $9\times 9$ block diagonal matrix shown in \cref{eq:tau}.

It is well-known that the IAM and the OOP methods suffer from the existence of spurious poles below the relevant thresholds in the scalar channels. A modification to the IAM was proposed in a later work~\cite{GomezNicola:2007qj} to remove this and also to ensure the existence of the Adler zero. We call this modified method mIAM-OOP or, simply, mOOP. The modification amounts to including an additional term in the inverse, which ensures that the spurious poles cancel out. It is given by
\begin{equation}
    T^{IJ}_m = T_2^{IJ} (T_2^{IJ} - T_{4P}^{IJ} - T_2^{IJ} G T_2^{IJ} + A^{IJ}_m)^{-1} T_2^{IJ}\,,
\end{equation}
where
\begin{equation}
    A^{IJ}_m = \left. T_{4P}^{IJ} \right\vert_{\sigma = \sigma_2} - \left( \left. \frac{d}{ds}(T_2^{IJ} - T_{4P}^{IJ})\right\vert_{\sigma = \sigma_2} \frac{(\sigma_2 - \sigma_A)(\sigma - \sigma_2)}{\sigma - \sigma_A} \right)\,.
\end{equation}
Here, $\sigma_2$ is the point such that $T_2^{IJ}(\sigma_2) = 0$, which is the Adler zero approximation at the leading order in CHPT, $\sigma_A$ is the point such that $T^{IJ}(\sigma_A) = 0$, which is the Adler zero in the full amplitude. For the spectator momenta and three-body invariant masses considered in the following analyses, the two-body invariant mass runs well below the threshold and even into the negative region. Therefore, this modification is crucial to ensure that no spurious imaginary parts are generated from the spectator momentum integration picking up unphysical sub-threshold poles, leading to unitarity violations. Furthermore, on-shell effects dominate in the triangle singularity region. Therefore, we do not stress on the dynamics of the two-body system in the negative invariant mass region, and use a damping factor in this region, if needed. This damping factor modification is given by
\begin{equation}
    T^{IJ}_m = e^{(\sigma - \sigma_0)/m_\pi^2} \left. T^{IJ}_m \right\vert_{\sigma=\sigma_0},\quad \sigma<\sigma_0,
\end{equation}
with $\sigma_0$ typically chosen as $0~\textrm{GeV}$, which also removes additional singularities well below the two-body threshold. The two-body amplitudes from OOP and mOOP are given in Fig.~\ref{fig:OOP}. This modification turns out not to be trivial in the case of coupled-channel systems and requires one to remove the spurious pole only along the null direction of the matrix. In the original work, the low-energy constants were fit to experimental data. We made no attempt at refitting these low-energy constants, but only ensured that the masses of the two-body resonances in the isobars agreed reasonably well with PDG by tuning the constants slightly. This is critical since the triangle singularity is very sensitive to the masses of two-body resonances, and particularly the $K^*(892)$. Since the other components of the IVU formalism are controlled by the formalism itself, we can proceed to solving the IVU equation.

\subsection{Solving the IVU Equations}
\label{sec:solveIVU}
We must first solve the IVU equation given in Eq.~\eqref{eq:LS-HB}. This is a Fredholm equation of the second kind, which is typically solved using the resolvent formalism. In order to solve this numerically, we discretize the integral and use Gaussian quadrature. In our case, the quantities are $9 \times 9$ matrices corresponding to the $9$ channels considered. Solving the IVU equation amounts to solving the matrix equation given by 
\begin{equation}
    T = (1 - (B + C)\tilde\tau)^{-1} (B + C)\,,
\end{equation}
where the inverse factor is the resolvent and the relevant Gaussian quadrature weights are absorbed into $\tilde\tau$. The resolvent is singular in certain kinematic regions, which makes the equation unsolvable directly for real kinematic variables. We resort to integration by deforming the integration contour into the complex plane. In that, we promote all the spectator momenta to complex variables, whereas the three-body invariant mass is still real. The complex integration contour, called the spectator momentum contour (SMC), is defined by~\cite{Sadasivan:2021emk} 
\begin{equation}
    f_{\textrm{SMC}}(t; V_0, w, \Lambda) =  t + i V_0 (1 - e^{-t/w}) (1 - e^{(t-\Lambda)/w}).
    \label{eq:SMC}
\end{equation}
\begin{figure}[t]
    \centering
    \includegraphics[width=0.99\linewidth]{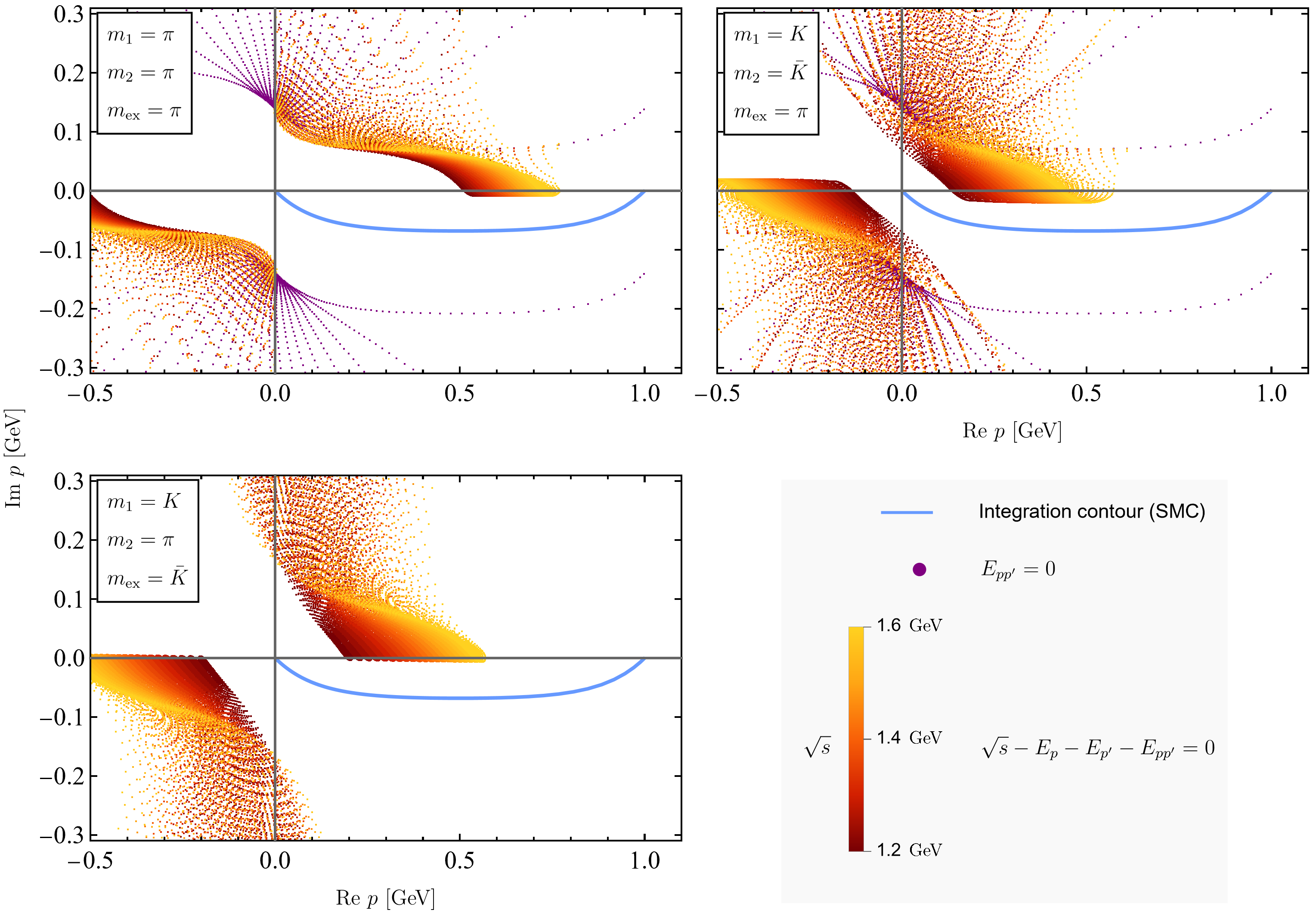}
    \caption{The singularities in the exchange term arising from Eq.~\eqref{eq:singularities}. These are handled by deforming the integration contour in the complex plane. Due to the conservation of strangeness, there are three different combinations of the spectators possible, which are given in the three figures. The singularities correspond to different three-body invariant masses, outgoing momenta values along the SMC, and $\cos\theta$ values in the interval $[-1,1]$.}
    \label{fig:singularitites}
\end{figure}%
In principle, assuming the isobar component is analytic, the singularities in the resolvent can arise only from the exchange term in \cref{eq:B-term}. Specifically, one can have singularities corresponding to
\begin{subequations}
    \begin{equation}
        E_{pp^\prime} = \sqrt{p^2 + p^{\prime2} + 2 p p^\prime\cos\theta + m_{\textrm{ex}}^2} = 0,
            \label{eq:singularities1}
    \end{equation}
    \begin{equation}
        \sqrt{s} - \sqrt{p^2 + m^2} - \sqrt{p^{\prime2} + m^{\prime2}} - E_{pp\prime} = 0,
    \end{equation}
    \label{eq:singularities}
\end{subequations}
where $m_{\textrm{ex}}$ denotes the mass of the exchanged particle, and unprimed and primed variables denote the incoming and the outgoing particles, respectively.

For the $9$ channels that we have, due to conservation of strangeness, we need to consider only three different combinations of the spectators. This is depicted in Fig.~\ref{fig:singularitites}. The figures show the singularities for various three-body invariant mass values, outgoing spectator momentum taking its values along the SMC, and $\cos\theta$ taking its value in the interval $[-1, 1]$. The SMC parameters are chosen as $V_0 = -0.5~\textrm{M}_\pi$, $w = 0.8~\textrm{M}_\pi$ and $\Lambda = 1~\textrm{GeV}$. Having solved the integral equation given in Eq.~\eqref{eq:LS-HB}, we need to integrate over the incoming spectator momentum as well, which is given by the quantity $\tilde\Gamma$ in \cref{eq:gammatilde_PW}. At this point, the amplitudes are given for an arbitrary three-body invariant mass and outgoing spectator momentum values along the SMC. In order to evaluate the amplitudes for real momenta, one needs to analytically continue the amplitudes evaluated for the discrete complex momenta back to the real axis. One choice for this task is the Padé approximant. However, we employ continued fractions here, using Thiele's interpolation formula, to construct the analytic continuation. For each fixed value of three-body energy $\sqrt{s}$, for example, the amplitude $\tilde\Gamma(s,q)$ is known numerically at the discrete points $q_i$ along the complex SMC. We construct the analytic continuation as
\begin{align}
    C_N(q) =
    \cfrac{\tilde{\Gamma}(\sqrt{s},q_1)}{
        1 + \cfrac{a_1(q-q_1)}{
            1 + \cfrac{a_2(q-q_2)}{
                1 + \cdots + \cfrac{a_{N-1}(q-q_{N-1})}{1}
            }
        }
    }\,.
\end{align}
The coefficients $a_i$ are obtained recursively by requiring that $C_N(q)$ matches the known input values at the points $q_i$, i.e.,
\begin{align}
    C_N(q_i)=\tilde{\Gamma}(\sqrt{s},q_i),\quad i=1,\dots,N.
\end{align}
We found that continued fractions better reproduce the triangle singularity compared to Padé approximants, at the cost of also introducing more spurious poles. For a more in depth discussion, see Ref.~\cite{Sakthivasan:2024uwd}. We ensured that there are no spurious poles in the region of interest. Padé approximants and continued fractions are both rational functions and are equivalent to one another at certain orders. The analytically continued $\tilde\Gamma$ can be evaluated for arbitrary three-body invariant masses and outgoing spectator momenta, with the caveat that one should be cautious with its region of validity.

The final amplitudes are given by adding the disconnected part $D\tau$ to $\tilde\Gamma$ and multiplying it by the outgoing spectator, which is exactly the quantity $\breve\Gamma$. At this point, one can integrate over the outgoing state phase space, which gives the effective decay width although the additional integration adds to the complexity of the problem. Alternatively, one can also compare the absolute squared values to the experimental intensities if the experimental analyses are carried out for a fixed kinematic window in the outgoing two-body invariant mass.  The latter is called the freed-isobar model in the COMPASS analysis~\cite{COMPASS:2015gxz}, which is discussed in the next section. 

\subsection{Experimental Data}
\label{sec:expdata}
The experimental data that we consider here to study the triangle singularity are the analyses of the $3\pi$ final state processes in the $p \pi^- \to p \pi^- \pi^+ \pi^-$ scattering through a pomeron exchange, with $t^\prime$ as the transferred momentum, see Ref.~\cite{COMPASS:2015gxz}. Specifically, we consider the experimental data for the freed-isobar model in the $(\pi^- \pi^+)_s \pi^-$ final state in a relative $P$-wave between the two-body cluster and the $\pi^-$, and the two-body cluster is in a relative $S$-wave, as denoted. This is shown in Fig.~\ref{fig:diffractive scattering}. The freed-isobar model allows one to carry out an analysis in bins of invariant masses in the two-body cluster, and we consider the bin $0.96 < m_{\pi^-\pi^+}/\textrm{GeV} < 1.00$, which includes the $f_0(980)$. Experimentally, this is the interval that shows an enhancement in the scattering amplitudes mimicking a resonance. The intensities used in our fits are given in App.~\ref{app:exp-data}, containing intensities for $12$ three-body invariant masses $s$ for four different transferred momentum $t^\prime$ bins. The intensities correspond to the absolute squared of the matrix elements for the production of a two-body cluster and a spectator from some initial source. Since experimentally one cannot ignore final state interactions, this corresponds to the absolute squared of the quantity $\breve\Gamma$ in the IVU formalism in the relevant channel, which would be channel $5$ containing the $f_0(500)/f_0(980)$ and $\pi$ with the isobar parameterized as a $\pi\pi$ state. It should be noted that the triangle singularity, which arises in a $K\bar{K}$ recombination, is present in channel $6$, which corresponds to a $K\bar{K}$ isobar state. However, the final isobar allows for in-flight transitions between the $\pi\pi$ and $K\bar{K}$ states, making the triangle singularity present in channel $5$ as well. The dependence of the intensities on the exchange momentum $t^\prime$ is not trivial, and the analysis is carried out in four different bins in $t^\prime$.
\begin{figure}[t]
    \centering
    \begin{tikzpicture}
        \begin{feynman}
            \vertex (pi_in) at (0, 0) {\(\pi^-\)};
            \vertex (p_in)  at (0, -4) {\(p\)}; 
            
            \vertex (v_top) at (2, -1);
            \vertex (v_bot) at (2, -3);
            
            \node [draw, circle, fill=black, minimum size=4.2mm, inner sep=0pt] (T) at (3.5, -1) {};
            
            \vertex (pi_out1) at (5.5, 0) {\(\pi^-\)};
            \vertex (pi_mid)  at (5.5, -1) {\(\pi^+\)}; 
            \vertex (pi_out2) at (5.5, -2) {\(\pi^-\)}; 
            \vertex (p_out)   at (5.5, -4) {\(p\)}; 
            
            \diagram* {
                (pi_in) -- [plain] (v_top) -- [plain] (T),
                (p_in) -- [fermion] (v_bot) -- [fermion] (p_out),
                (v_top) -- [double, thick, edge label=\(\mathbb{P}\)] (v_bot),
    
                (T) -- [plain] (pi_out1),
                (T) -- [plain] (pi_mid),
                (T) -- [plain] (pi_out2),
            };
            
            \node at (-0.2, -2) {$\bar{s}\to$};
            \node at (2.75, -0.7) {$X^-$};
            \node at (2.75, -1.2) {$s\to$};
            
            \node [above=0.2cm of v_top] {\(\downarrow t^\prime\)};
            
            \draw [decorate, decoration={brace, amplitude=5pt}, thick] (5.9, -0.8) -- (5.9, -2.2) node [midway, right=0.2cm] {Freed-isobar};
    
            \draw [dashed, blue, thick] (3.1, 0.5) rectangle (8.1, -2.5);
        \end{feynman}
    \end{tikzpicture}
    \caption[Diffractive production of $p\pi^-\to p\pi^-\pi^-\pi^+$]{Diffractive production of a three-pion system in $\pi^- p$ scattering via Pomeron exchange, adapted from Ref.~\cite{COMPASS:2018uzl}, where the pomeron represents an effective colorless exchange with vacuum quantum numbers. The intermediate state $X^-$ decays into $\pi^-\pi^-\pi^+$ through an isobar $\xi^0$ and a bachelor $\pi^-$ (we call it spectator in our projects), with relative orbital angular momentum $L$. The blue-dashed box shows the amplitudes corresponding to the experimental data which is matched to our formalism. Non-resonant effects such as the Deck effect are not shown here.
    }
    \label{fig:diffractive scattering}
\end{figure}

The IVU equation has degrees of freedom in the dissociation vertices $D$ and also the contact term $C$. The experimental amplitudes factorize into (a) an amplitude describing the production of $X^-$, see \cref{fig:diffractive scattering}, and (b) an amplitude describing its decay into the $3\pi$ state. The IVU formalism explicitly handles the production of the $3\pi$ state from some source, and the source itself is parametrized through the fit parameters. The contact term may also contain poles that are to be fit to the data, appearing both in $C$ and $D$, see \cref{eq:C} and \cref{eq:D}. We test the inclusion of poles not to mimic the triangle singularity, but to study the effect of the close-by $a_1(1260)$. In simpler fits without explicit resonance, we only fit the dissociation vertices $d_i$ of \cref{eq:D}. Therefore, we consider the data corresponding to the four different exchange momenta to be independent and obtain four different fits. The pole parameters in the IVU formalism are independent of the transferred momenta, which means that these parameters are shared across the four different analyses in the different $t'$ bins. It is also important to note that there are also non-resonant background effects like the Deck effect that contribute to the amplitude. These affect the pole position of any actual pole, if present. A very essential information is that the experimental data for the freed-isobar model include only statistical errors and no systematic errors. Due to the extremely large number of statistics in the experiment, the statistical errors are tiny. The analysis also claims that systematic errors are the dominant source of errors in the data. Since these are not available, the obtained fits might be in ostensible visual agreement with the data with poor goodness of fit test agreements. This is extensively discussed in Sec.~\ref{sec:Results}.

\section{Results}
\label{sec:Results}
\subsection{The Triangle Singularity}
Before we tackle the challenging task of fitting the full $9$ coupled-channel amplitudes with $80$ Gauss nodes per channel to the experimental data, we consider something simpler. In a previous work~\cite{Sakthivasan:2024uwd}, it was shown how final state interactions affect the triangle singularity in an $a_1(1420)$-like toy-model. In this section, we show how the triangle singularity manifests itself in a system with the exact quantum numbers of the mesons involved. 
\begin{figure}
    \centering
    \includegraphics[width=0.45\linewidth]{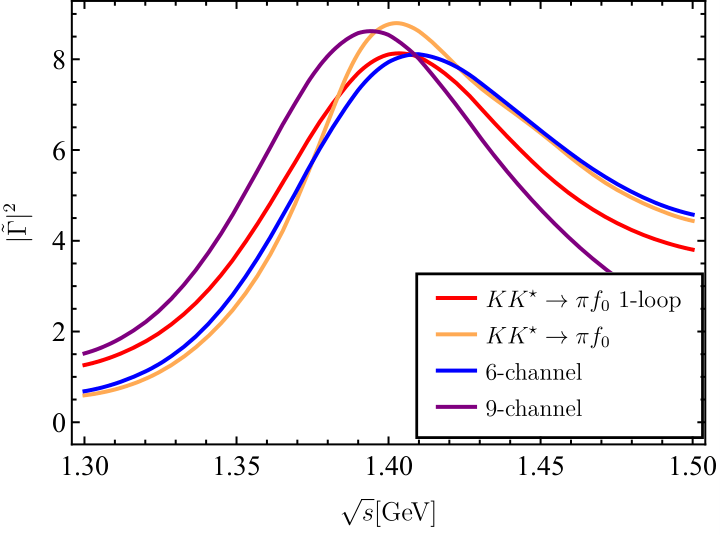}
    \caption{ Comparison of normalized $|\tilde\Gamma|^2$ for different situation at $\sqrt{\sigma}=0.99 $  GeV: the one-loop $K K^\star \to \pi f_0$ contribution (no coupled-channel rescattering, at all), the $K K^\star \to \pi f_0$ amplitude including rescattering effects (four-channel model), the six-channel model, and the full nine-channel model. All curves are normalized to the same area beneath them to make them comparable.
    }
    \label{fig:GamtilNorm}
\end{figure}%
First, in \cref{fig:GamtilNorm} we assess the influence of the different ingredients of the amplitude by setting all resonance couplings $g_j=0$, $d_i=M_\pi$ for vector isobar channels ($i\in\{1,\dots, 4\}$), and $d_i=1$ for scalar isobar channels ($i\in\{5,\dots, 9\}$) in \cref{eq:C} and \cref{eq:D}. In other words, there is no contact term, no resonance, and the production amplitude of all channels is chosen to be of similar magnitude. In addition, all curves in \cref{fig:GamtilNorm} are normalized to the same area below them to facilitate comparability -- the overall normalization is arbitrary anyway. The invariant mass is fixed at $\sqrt{\sigma}=0.99$~GeV where the triangle signal is the strongest. The red curve shows the situation when $\tilde\Gamma$ in \cref{eq:gammatilde_PW} reduces to a $K^*\bar K K$ one-loop triangle, with a final $K\bar K\to\pi\pi$ transition in the $f_0$ isobar. This corresponds to the result discussed primarily in the literature~\cite{Mikhasenko:2015vca, Mikhasenko:2015oxp, Aceti:2016yeb, Guo:2019twa, Sakthivasan:2024uwd, COMPASS:2020yhb}, even though often the final, non-diagonal isobar transition is not included. It should be emphasized that here the explicit $f_0(980)$ two-channel amplitude leads to a substantial enhancement of the triangle singularity in the three-pion spectrum: first, the $f_0$ mass coincides with the TS condition in $\sqrt{\sigma}$, and, second, the $f_0$ provides a strong $K\bar K\to\pi\pi$ transition to the observed pion channel.

The orange curve shows the model including only the channels 3, 4, 5 and 6. This is the minimal unitary model with final state interaction containing the $K^*\bar K$ structure needed for the triangle and the $\pi f_0$ channels needed for populating the final $3\pi$ state. One sees a moderate change in the shape of the triangle singularity. The blue curve shows the six-channel case, which includes, in addition, the two $\pi\rho$ channels 1 and 2 that are phenomenologically important for the $a_1(1260)$. They lead to another small modification. The purple curve shows the nine-channel model, which is the full production amplitude and adds the repulsive $\pi\pi$ and $\pi K$ isobars, as well as the $\kappa$ isobar channels. These ingredients move the position of the peak by around 20 MeV and make it a bit narrower. Overall, the persistence of this enhancement from the one-loop contribution to the full nine-channel calculation supports the interpretation that the observed structure is primarily driven by the dynamics of the triangle singularity mechanism at one loop. This finding is in line with the semi-quantitative model of Ref.~\cite{Sakthivasan:2024uwd}.
\begin{figure}
    \centering
    \includegraphics[width=0.98\linewidth]{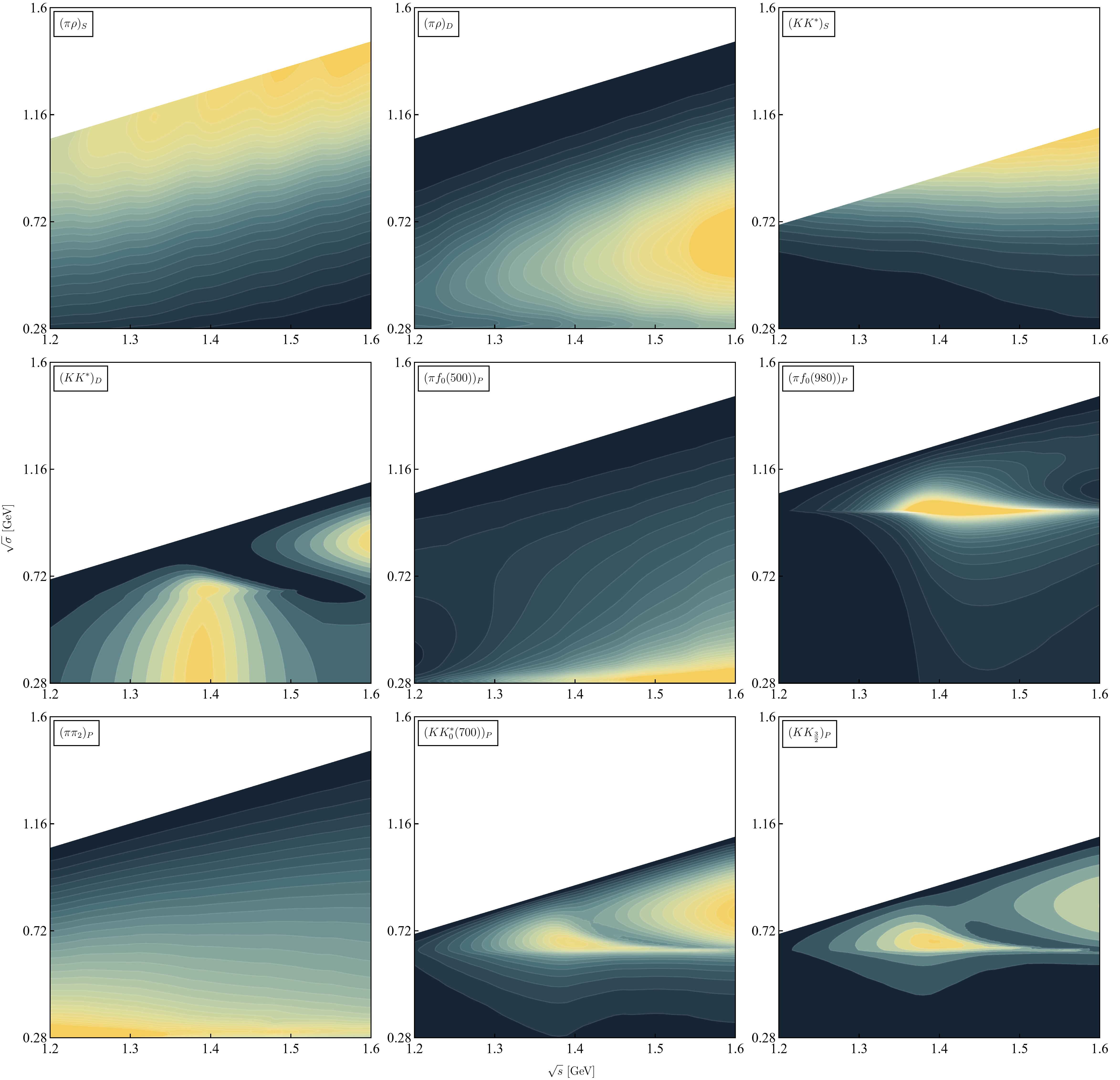}
    \caption{Contour plots of $|\tilde{\Gamma}|^2$ of \cref{eq:gammatilde_PW} for the nine different final states as functions of three-body energy $\sqrt{s}$ and isobar sub-energy $\sqrt{\sigma}$. Lighter (darker) colors correspond to larger (smaller) values of $|\tilde{\Gamma}|^2$. A pronounced localized enhancement is observed in the $(\pi f_0(980))_P$ channel, consistent with the expected triangle singularity, but other channels also show enhancements as discussed in the main text.}
    \label{fig:Gamma-tilde contour}
\end{figure}%
Next, we discuss the production amplitude for all nine channels as shown in \cref{fig:Gamma-tilde contour} (with $g_j=0$ and $d_i$ still set to the above-mentioned values). Among these channels, the most prominent localized enhancement appears in the $(\pi f_0(980))_P$ channel. This is the channel where the triangle singularity is expected to contribute, corresponding to the kinematic configuration in which the intermediate $K$, $\bar K$, and $K^\star$ can approach the on-shell condition. 
In addition, we also observe the $\pi K$ $S$-wave threshold in channels $8$ and $9$. 

Back to the manifestation of the triangle singularity, we consider the solutions when only the $K^*(892)K$ in relative $S$- and $D$-waves (the channels $3$ and $4$) and the $f_0(500)/f_0(980)\pi$ with the isobars parametrized as a $\pi\pi$ or a $K\bar{K}$ state (the channels $5$ and $6$) channels are turned on. This corresponds to the orange line in \cref{fig:GamtilNorm}.
\begin{figure}[t]
    \centering
    \begin{minipage}[b]{0.4\linewidth}
        \begin{subfigure}[b]{\linewidth}
            \centering
            \includegraphics[width=\linewidth]{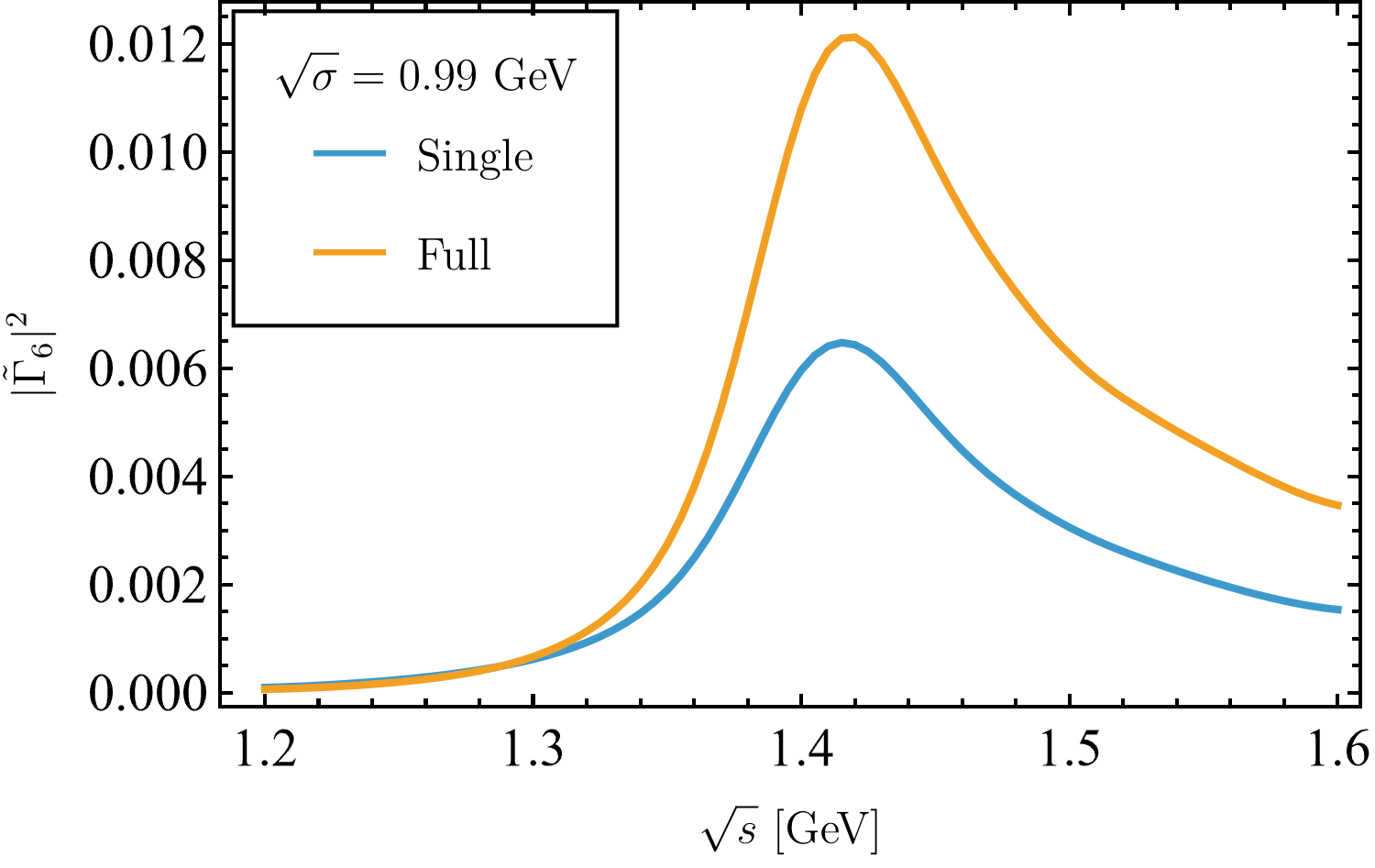}
            \caption{}
            \label{fig:TS-LS}
        \end{subfigure}\\
        \vspace{4ex}
        \begin{subfigure}[b]{\linewidth}
            \centering
            \includegraphics[width=\linewidth]{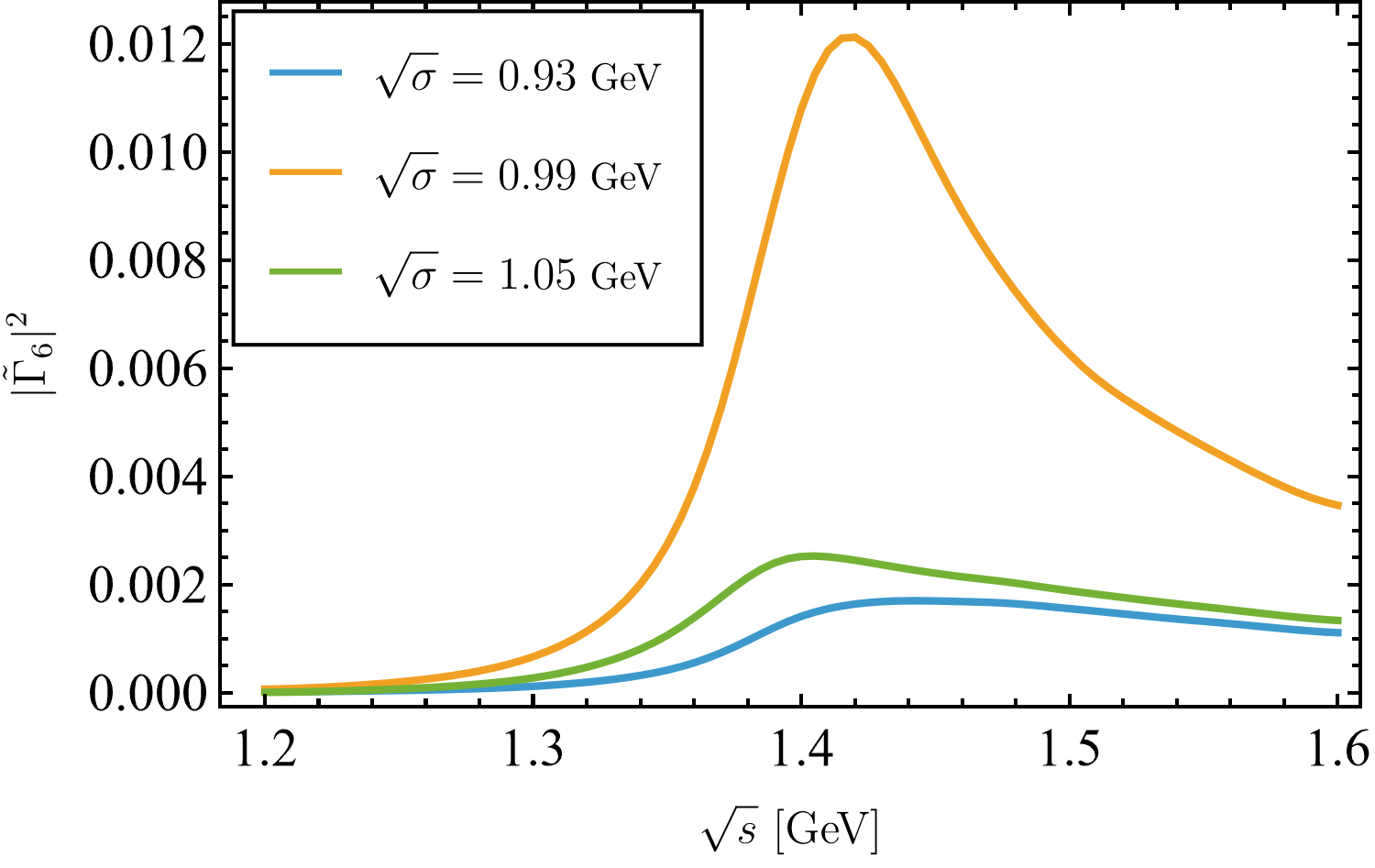}
            \caption{}
            \label{fig:TS-sigma}
        \end{subfigure}
    \end{minipage}%
    \hspace{4ex}
    \begin{minipage}[b]{0.52\linewidth}
        \begin{subfigure}[b]{\linewidth}
            \centering
            \includegraphics[width=\linewidth]{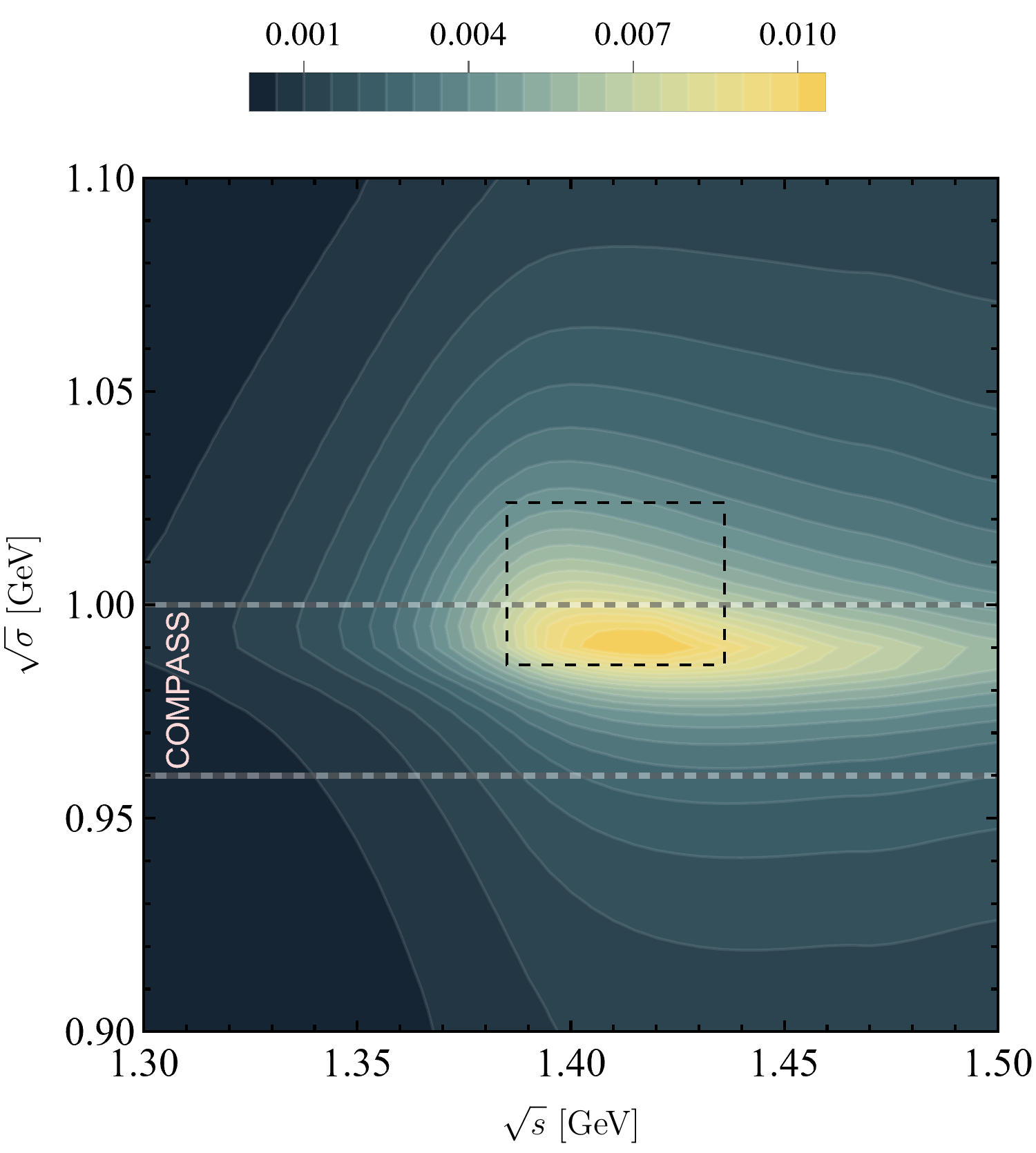}
            \caption{}
            \label{fig:TS-contour}
        \end{subfigure}
    \end{minipage}
    \caption{The triangle singularity when only the $K^*(892)K$ and the $f_0(500)/f_0(980)\pi$ channel are turned on. (\subref{fig:TS-LS})~the triangle singularity for a fixed $\sqrt{\sigma}$ with a single rescattering and the full amplitude after final state interactions are taken into account; (\subref{fig:TS-sigma})~the triangle singularity for three different $\sqrt{\sigma}$ after final state interactions are taken into account; (\subref{fig:TS-contour})~the kinematic region on the $\sqrt{s}\sqrt{\sigma}$-plane in which the triangle singularity is present, where the dashed box denotes the theoretical singular region, and the gridlines corresponds to the experimental bin interval~\cite{COMPASS:2015gxz}.}
    \label{fig:TS}
\end{figure}%
In Fig.~\ref{fig:TS}, we plot the square of the absolute values of $\tilde\Gamma$ in the outgoing channel $6$ corresponding to a $K \bar{K}$ recombination. This was exactly the quantity that was studied in Ref.~\cite{Sakthivasan:2024uwd}, thus enabling direct comparisons with the results presented there. Figs.~\ref{fig:TS-LS} and~\ref{fig:TS-sigma} show the amplitude as a function of the three-body invariant mass $\sqrt{s}$ and for a given outgoing two-body invariant mass $\sqrt{\sigma}$. Fig.~\ref{fig:TS-LS} shows two different curves: the curve labeled ``single'' corresponding to a single rescattering ($T = B$), and the curve labeled ``full'' corresponding to the full solution. For the curves shown in this figure, the two-body invariant mass $\sqrt{\sigma}$ is fixed at $0.99~\textrm{GeV}$, which corresponds to the PDG value for the mass of the $f_0(980)$. This figure suggests a significant contribution from the final state interactions compared to the toy-model, see Ref.~\cite{Sakthivasan:2024uwd}. However, the location of the peak remains unchanged when the final state interactions are taken into account, as expected from that reference. As the overall normalization is arbitrary, anyway, this almost multiplicative change through the final state interaction would be unobservable in practice. Fig.~\ref{fig:TS-sigma} shows three different curves corresponding to three different values of $\sqrt{\sigma}$. The figure shows that the triangle singularity is pronounced when $\sqrt{\sigma}$ corresponds to the mass of the $f_0(980)$, which supports the claims made with the toy-model in Ref.~\cite{Sakthivasan:2024uwd}. Finally, Fig.~\ref{fig:TS-contour} shows the amplitudes on the $\sqrt{s}-\sqrt{\sigma}$ plane, similar to \cref{fig:Gamma-tilde contour}. The triangle singularity occurs only in a restricted interval in $\sqrt{s}$ and $\sqrt{\sigma}$ for an outgoing spectator pion, see Ref.~\cite{Sakthivasan:2024uwd}. This is given by the dashed box in the figure. This result follows from the solutions to the Landau equations, which are derived under the assumption that the particles involved in the rescattering process are stable and have zero widths. Outside this kinematic region where the classical Coleman-Norton~\cite{Coleman:1965xm} configuration can occur, the Landau equations cannot be solved for real incoming and outgoing invariant masses. Further, in reality, the particles involved in the rescattering processes have finite widths whose effects on the triangle singularity are non-trivial. These effects manifest as a smearing out of the otherwise logarithmic singularity. This is further exacerbated by the rather narrow $f_0(980)$ making the triangle singularity more prominent in the observed lineshapes. Additionally, the figure also shows the interval labeled ``COMPASS'' corresponding to the experimental bin in which the analysis of the COMPASS experiment is carried out for the data used in our work, as discussed in Sec.~\ref{sec:expdata}.
\begin{figure}[t]
    \centering
    \includegraphics[width=0.98\linewidth]{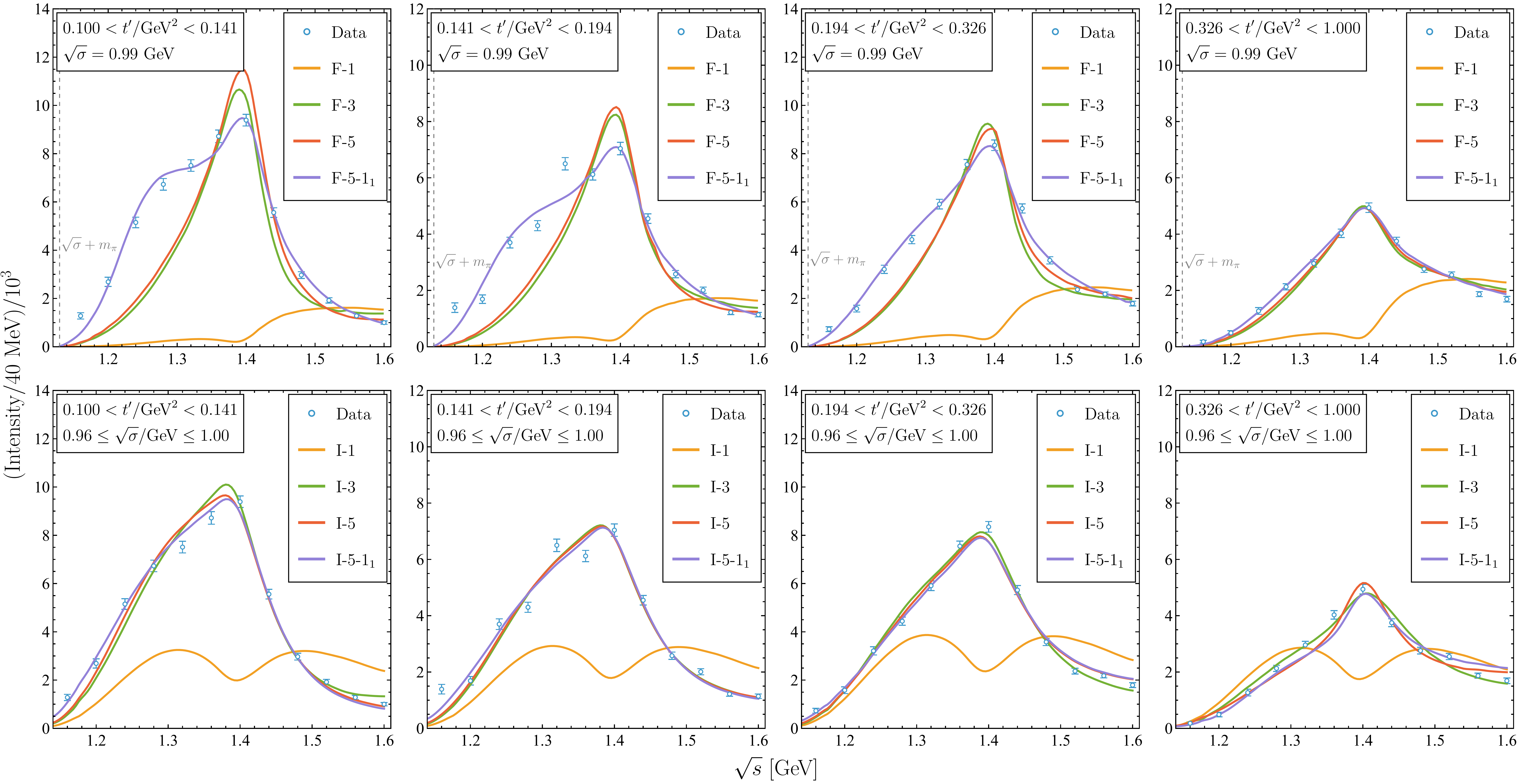}
    \caption{In the top row, independent fits for various momentum transfer $t^\prime$ intervals for a fixed outgoing two body invariant mass $\sqrt{\sigma} = 0.99~\textrm{GeV}$. In the bottom row, independent fits for various momentum transfer $t^\prime$ intervals with the outgoing two body invariant mass $\sqrt{\sigma}$ integrated over the interval $[0.96~\textrm{GeV}, 1.00~\textrm{GeV}]$. In each plot, the fit parameters corresponding to the curves are: (1) a single parameter for all channels given by $d_0$, (2) independent parameters in channels $3$ and $6$ given by $d_3$ and $d_6$, (3) independent parameters in channels $3$, $5$, $6$, $7$ and $8$ given by $d_3$, $d_5$, $d_6$, $d_7$ and $d_8$, and (4) same as before but a pole that couples to channel $1$ with parameters $g_1$ and $m_{\textrm{fit}}$ is independently fit for every $t^\prime$. In the top row, the dashed gray line on the left shows the isobar-spectator threshold. There is no single isobar-spectator threshold for the bottom row. The COMPASS experimental data shown here~\cite{COMPASS:2015gxz} are used as discussed in the main text.}
    \label{fig:first-order-fits-FI}
\end{figure}%

\subsection{Fits to Experimental Data}
\label{sec:Fits}
Coming back to the full formalism which respects three-body unitarity in all $9$ coupled-channels, the free parameters are contained in the dissociation vertices $D$ and contact terms $C$ corresponding to different production strengths in different channels, and, possibly, a resonance with different channel couplings. This freedom allows us to fit the amplitudes to the experimental data and controls the strength of the various channels. We do not include analytic contact terms in \cref{eq:C} in addition to the possible pole term, and only constant $d_i$ from \cref{eq:D} due to scarcity of data. The primary reason for ignoring possible energy and momentum dependence in C and D is to see whether one can reproduce the peak observed in the data with a minimal set of parameters.
\begin{figure}[t]
    \centering
    \begin{minipage}{0.5\linewidth}
        \centering
        \includegraphics[width=0.9\textwidth]{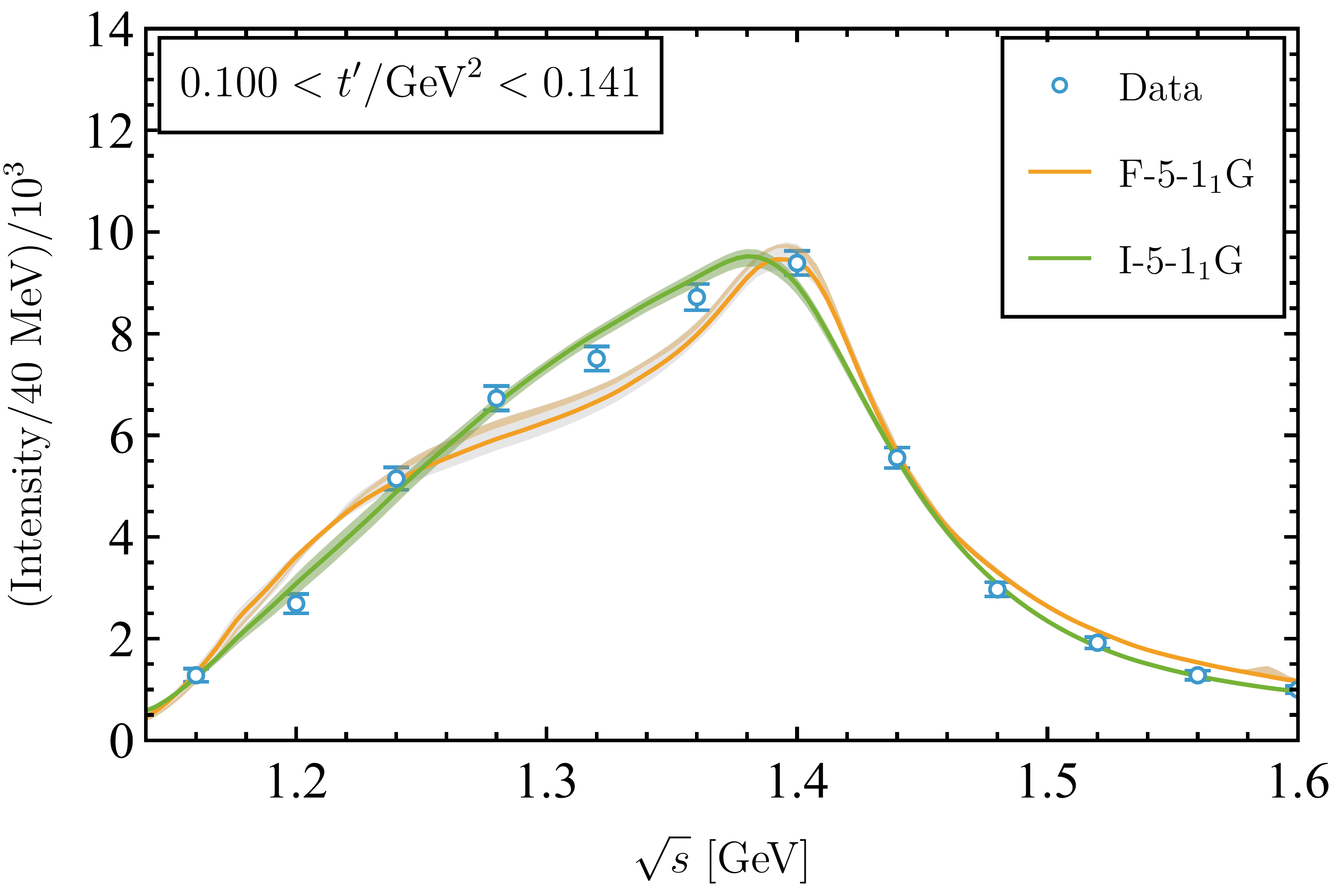}
    \end{minipage}%
    \begin{minipage}{0.5\linewidth}
        \centering
        \includegraphics[width=0.9\textwidth]{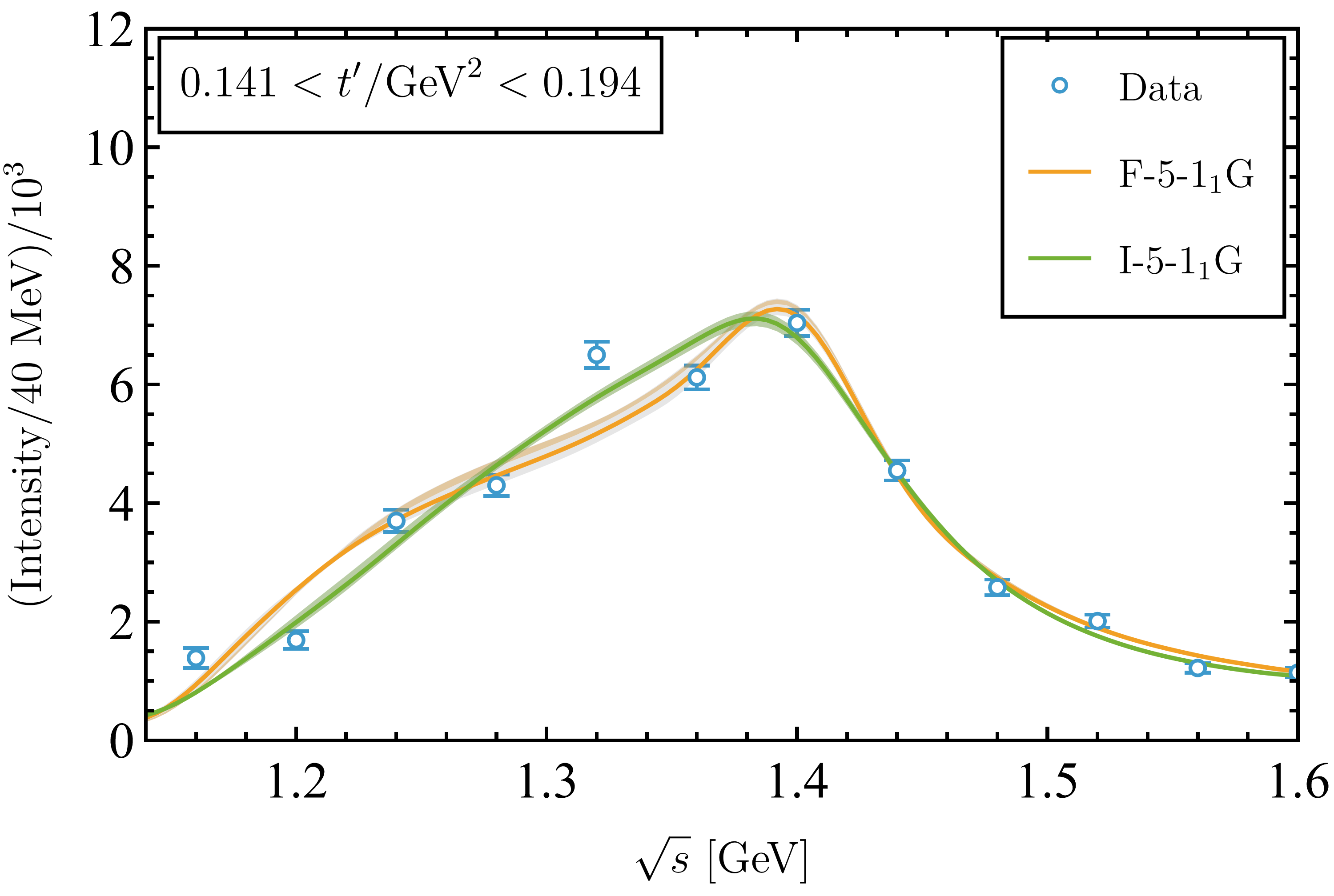}
    \end{minipage}\vspace{1em}
    \begin{minipage}{0.5\linewidth}
        \centering
        \includegraphics[width=0.9\textwidth]{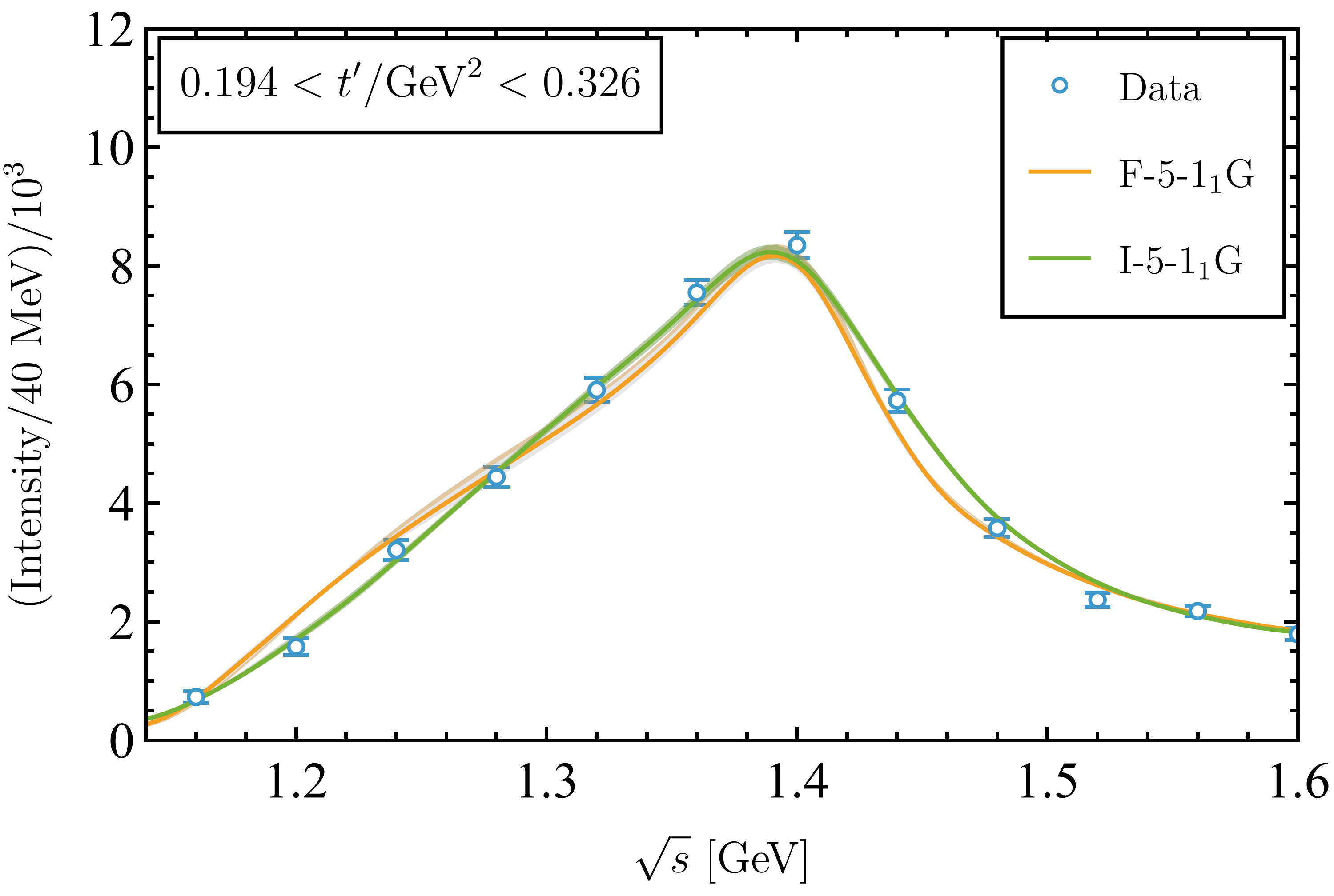}
    \end{minipage}%
    \begin{minipage}{0.5\linewidth}
        \centering
        \includegraphics[width=0.9\textwidth]{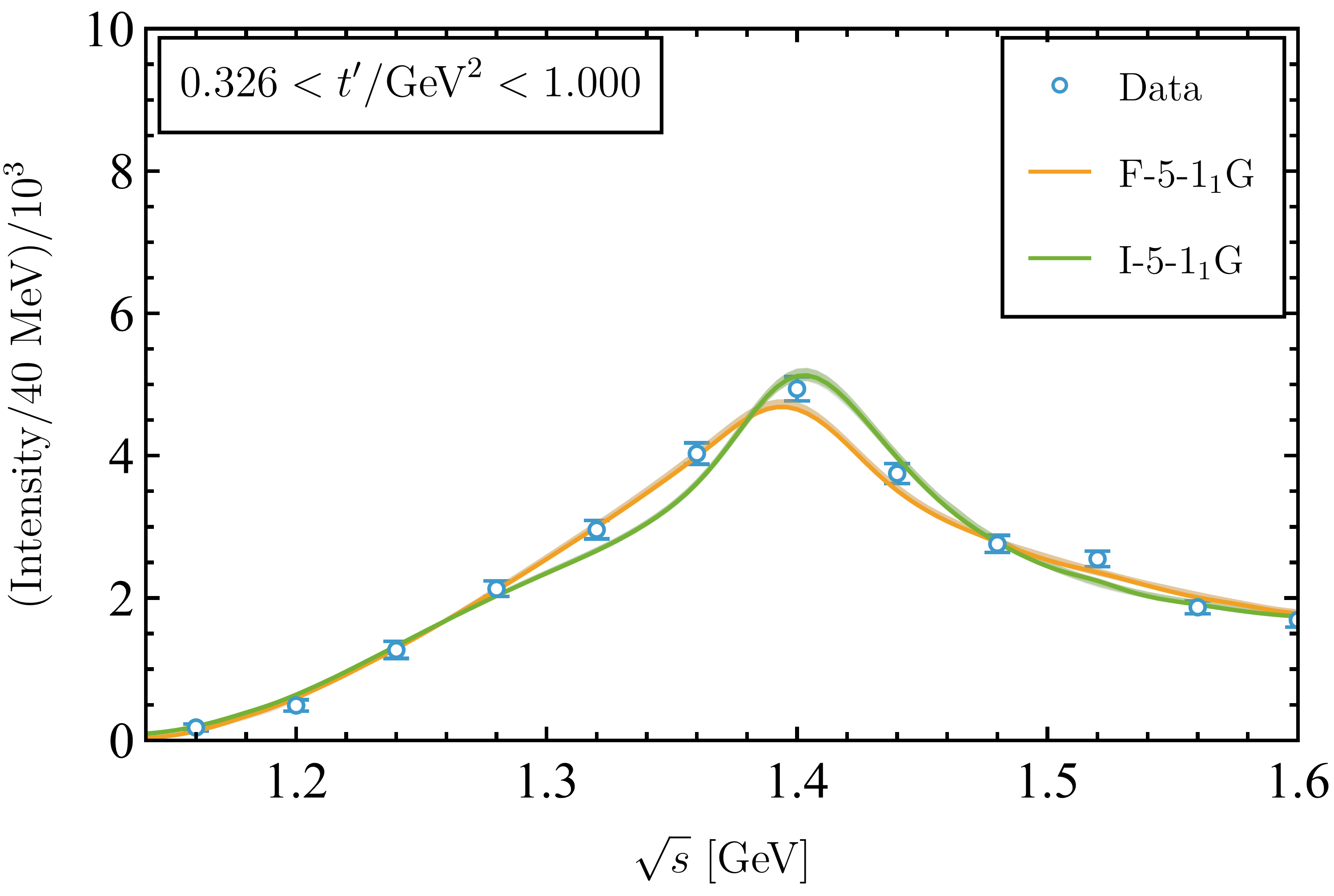}
    \end{minipage}
    \caption{The overall fits for various momentum transferred $t^\prime$ intervals. In the fits, the channels $3$, $5$, $6$, $7$ and $8$ get independent parameters. Additionally, a pole with its bare mass and couplings held the same across the plots is also fit. The first curve (F-5-1\textsubscript{1}G) in each plot shows the case when the outgoing two-body invariant mass $\sqrt{\sigma}$ is fixed at $0.99~\textrm{GeV}$, whereas the second curve (I-5-1\textsubscript{1}G) in each plot corresponds to an integration over the interval $[0.96~\textrm{GeV}, 1.00~\textrm{GeV}]$ in $\sqrt{\sigma}$. The bands show the statistical and the total quadratic errors in the fits as described in the main text. Note that the total quadratic errors are almost exactly equal to the statistical errors since the systematic errors are negligible.  The COMPASS experimental data shown here~\cite{COMPASS:2015gxz} are used as discussed in the main text.}
    \label{fig:full-fits}
\end{figure}%

In our fits, we consider the square of the absolute value of $\Breve{\Gamma}$ in the $JLS$ basis given by
\begin{equation}
    \vert \breve\Gamma_5 \vert^2 = \lvert \tau_{55} (\tilde\Gamma + D)_5 + \tau_{56} (\tilde\Gamma + D)_6 \rvert^2\,,
\end{equation}
where the two terms emphasize that the isoscalar $\pi\pi$ state (channel $5$) can be populated through the diagonal isobar transition $\tau_{55}$, but more importantly, also through the non-diagonal $K\bar K\to\pi\pi$ transition $\tau_{56}$ that follows the triangle loop. Furthermore, we turn on all $9$ channels available to us in our fits, and carry out independent fits for every momentum transfer $t^\prime$. To understand how the different channels available to us affect the amplitudes, we consider various models involving different numbers of degrees of freedom. As mentioned before, the outgoing isobar invariant mass is taken to be a bin of width $40~\textrm{MeV}$ experimentally. To account for this, we consider primarily two different models: 1) models labeled $\textrm{F}$ where the outgoing invariant mass $\sqrt{\sigma}$ is fixed at $0.99~\textrm{GeV}$, which corresponds to the mass of the $f_0(980)$, i.e.,
\begin{equation}
    \vert \breve\Gamma_5 \vert^2_\textrm{F} = \left. \lvert \tau_{55} (\tilde\Gamma + D)_5 + \tau_{56} (\tilde\Gamma + D)_6 \rvert^2 \right\vert_{\sqrt{\sigma} = 0.99~\textrm{GeV}}\,,
    \label{eq:model-F}
\end{equation}
and 2) models labeled $\textrm{I}$ where the outgoing invariant mass is integrated over the experimental interval, i.e.,
\begin{equation}
    \vert \breve\Gamma_5 \vert^2_\textrm{I} = \int_{0.96~\textrm{GeV}}^{1.00~\textrm{GeV}} d\sqrt{\sigma}\ \lvert \tau_{55} (\tilde\Gamma + D)_5 + \tau_{56} (\tilde\Gamma + D)_6 \rvert^2\,.
    \label{eq:model-I}
\end{equation}
The latter conforms better to the experimental situation, since we cannot differentiate between the production of isobars with different invariant masses within this interval. However, the former shows the behavior of the amplitudes in the vicinity of the PDG value for the mass of the $f_0(980)$, allowing one to see how the non-resonant contributions from the isobar affect the amplitudes. The results of the fits corresponding to a fixed outgoing two-body invariant mass are given in the plots of the top row in Fig.~\ref{fig:first-order-fits-FI}, whereas the results of the fits when the outgoing two-body invariant mass is integrated over the experimental bin interval is given in the plots on the bottom row in Fig.~\ref{fig:first-order-fits-FI}. In both cases, we sequentially increase the number of fit parameters, which is given by the number following the label. We recall that the triangle singularity arises in a $K^*(892) K$ rescattering through a $K$ exchange, which corresponds to the channels $3$ and $4$ in our case. A transition from the kaon to the pion sectors is allowed only through the channels $5$ and $6$. Therefore, we expect the effect of the fit parameters corresponding to these channels to be the most significant. We also consider including a pole, which is given by the next number in the labels. In principle, the pole can couple to multiple channels. The number of channels to which a given pole couples is given by the subscript in the pole label. The $a_1(1260)$, for example, which lies quite close to the triangle singularity, has been observed to couple to both the $\rho(770)\pi$ and the $f_0(980)\pi$ channels, although predominantly to the former. 

Finally, we expect the pole fit parameters to be independent of the transferred momentum $t^\prime$, which means that we can consider an overall fit in which the pole parameters are shared. Such a fit is labeled by a \textit{G}. We recall that we still have to fix the parameters in the taming factor $H$ and the Blatt-Weisskopf factor. The latter is fixed as $\lambda_{BW} = 1/22~\textrm{M}_\pi^{-1}$, see \cref{eq:BWfactor}, so that we avoid any possible singularities, whereas the former is fixed so that the convergence of the amplitudes is ensured. To this end, $\lambda_H$, see \cref{eq:tamingfactor}, is chosen to be $0.4~\textrm{GeV}$ for the models labeled F, see \cref{eq:model-F}, and $0.3~\textrm{GeV}$ for the models labeled I, see \cref{eq:model-I}.
\begin{table}[t]
    \setlength{\tabcolsep}{6pt}
    \centering
    \caption{Main fit results for the available momentum transferred bins $t^\prime/\textrm{GeV}^2$: $(0.100, 0.141)$, $(0.141, 0.194)$, $(0.194, 0.326)$ and $(0.326, 1.000)$. The models labeled \textit{F} consider the outgoing two-body invariant mass to be fixed at $\sqrt{\sigma} = 0.99~\textrm{GeV}$, whereas the models labeled \textit{I} integrate over the interval $[0.96~\textrm{GeV}, 1.00~\textrm{GeV}]$ in the outgoing two-body invariant mass, see Eqs.~\eqref{eq:model-F}~and~\eqref{eq:model-I}. The number following that denotes the number of independent fit parameters used with $d_0$ and $d_i$ corresponding to an overall normalization and a fit parameter in channel $i$, respectively. The final number denotes the number poles, with the subscript denoting the number of couplings. In the absence of a pole, the parameters are fit independently to a given momentum transferred bin $t^\prime$. In the presence of poles, we can also consider an overall fit, denoted by ``G''. We provide both $\chi^2/\textrm{dof}$ and the $\tilde\chi^2/\textrm{dof}$ values, assuming systematic uncertainty as described in the main text.}
    \label{tab:fit-values}
    \begin{adjustbox}{max width=\linewidth}
        \begin{tabular}{|l l c | c c c c | c c c c|}
            \hline
            \multirow{2}{*}{Fit model} & \multirow{2}{*}{Parameters} & \multirow{2}{*}{dof} & \multicolumn{4}{c|}{$\chi^2$/dof} & \multicolumn{4}{c|}{$\tilde\chi^2$/dof} \\ \cline{4-11}
                &   &   & $t^\prime_1$ & $t^\prime_2$ & $t^\prime_3$ & $t^\prime_4$ & $t^\prime_1$ & $t^\prime_2$ & $t^\prime_3$ & $t^\prime_4$ \\ 
            \hline
            \hline
            F-1 & $\{d_0\}^{t^\prime}$ & $11$ & $505.22$ & $372.40$ & $434.05$ & $215.68$ & $126.30$ & $93.10$ & $108.51$ & $53.92$ \\ 
            \hline
            F-3 & $\{d_0, d_3, d_6\}^{t^\prime}$ & $9$ & $88.61$ & $57.75$ & $35.56$ & $5.48$ & $22.15$ & $14.44$ & $8.89$ & $1.37$ \\ 
            \hline
            F-5 & $\{d_0, d_3, d_5, d_6, d_8\}^{t^\prime}$ & $7$ & $99.91$ & $67.45$ & $37.11$ & $3.20$ & $24.98$ & $16.86$ & $9.28$ & $0.80$ \\ 
            \hline
            F-5-1\textsubscript{1}  & $\{d_0, d_3, d_5, d_6, d_8, d, g_1, m_\textrm{fit}\}^{t^\prime}$ & $4$ & $9.45$ & $13.43$ & $5.22$ & $2.65$ & $2.36$ & $3.36$ & $1.30$ & $0.66$ \\ 
            \hline
            \textbf{F-5-1\textsubscript{1}G} & $\{d_0, d_3, d_5, d_6, d_8, d\}^{t^\prime}, g_1, m_\textrm{fit}$ & $22$ & \multicolumn{4}{c|}{$\mathbf{9.34}$} & \multicolumn{4}{c|}{$\mathbf{2.34}$} \\ 
            \hline
            F-5-1\textsubscript{2}G & $\{d_0, d_3, d_5, d_6, d_8, d\}^{t^\prime}, g_1, g_5, m_\textrm{fit}$ & $21$ & \multicolumn{4}{c|}{$9.79$} & \multicolumn{4}{c|}{$2.45$} \\
            \hline
            \hline
            I-1 & $\{d_0\}^{t^\prime}$ & $11$ & $295.06$ & $164.82$ & $163.42$ & $61.69$ & $75.62$ & $43.14$ & $40.86$ & $15.42$ \\ 
            \hline
            I-3 & $\{d_0, d_3, d_6\}^{t^\prime}$ & $9$ & $10.79$ & $7.95$ & $2.57$ & $3.18$ & $2.70$ & $1.99$ & $0.64$ & $0.79$ \\ 
            \hline
            I-5 & $\{d_0, d_3, d_5, d_6, d_8\}^{t^\prime}$ & $7$ & $7.53$ & $8.39$ & $4.46$ & $6.64$ & $1.88$ & $2.10$ & $1.11$ & $1.66$ \\ 
            \hline
            I-5-1\textsubscript{1} & $\{d_0, d_3, d_5, d_6, d_8, d, g_1, m_\textrm{fit}\}^{t^\prime}$ & $4$ & $5.60$ & $11.39$ & $7.49$ & $15.72$ & $1.40$ & $2.85$ & $1.87$ & $3.93$ \\ 
            \hline
            \textbf{I-5-1\textsubscript{1}G} & $\{d_0, d_3, d_5, d_6, d_8, d\}^{t^\prime}, g_1, m_\textrm{fit}$ & $22$ & \multicolumn{4}{c|}{$\mathbf{4.75}$} & \multicolumn{4}{c|}{$\mathbf{1.19}$} \\ 
            \hline
            I-5-1\textsubscript{2}G & $\{d_0, d_3, d_5, d_6, d_8, d\}^{t^\prime}, g_1, g_5, m_\textrm{fit}$ & $21$ & \multicolumn{4}{c|}{$4.98$} & \multicolumn{4}{c|}{$1.25$}\\
            \hline
        \end{tabular}
    \end{adjustbox}
\end{table}%
In all our fits, we use the standard $\chi^2$ given by
\begin{equation}
    \chi^2 = \sum_{i} \frac{(\vert \breve\Gamma(\sqrt{s_i}) \vert^2_5 - \mathcal{I}_i)^2}{\sigma_i^2}\,,
\end{equation}
for testing the goodness of fit. The quantity $\mathcal{I}$ correspond to the experimental intensities from the COMPASS analysis -- specifically the intensities corresponding to the production of a pion and a two pion state in the interval $[0.96~\textrm{GeV}, 1.00~\textrm{GeV}]$ in the freed-isobar model with the same quantum numbers as the $a_1(1260)$, see Ref.~\cite{COMPASS:2015gxz} and also the discussion in Sec.~\ref{sec:expdata}. The various models, along with their corresponding $\chi^2$ per degree of freedom values, are summarized in Tab.~\ref{tab:fit-values}. The number of degrees of freedom in the fit is given by the difference between the number of data points ($12$ per momentum transferred $t^\prime$) and the number of fit parameters. We emphasize again that the absence of systematic errors in the data lead to rather large $\chi^2$ values per degree of freedom, indicating an underestimation of overall errors in the data. In all the cases, it can be seen that a single parameter corresponding to the overall normalization, which implies an equal dissociation into all $9$ channels, is not capable of reproducing the observed data. Interestingly, this differs from the analysis carried out in Ref.~\cite{COMPASS:2020yhb}. Here, a single triangle loop involving an on-shell $K^*(892)$, $K$ and $\bar{K}$ is taken to contribute to the amplitude, to which a background is added. This is then fit to the data. In our approach, the final-state interactions in the amplitudes are taken into account, and the background corresponds to the free propagation of the isobars. This approach is more theoretically sound, and it turns out that in our case the non-interacting contribution almost exactly cancels out with the interacting contribution leading to an accidental dip visible in the curves labeled as F-1 and I-1 in \cref{fig:first-order-fits-FI}.

Introducing independent fit parameters $d_3,\,d_6$ in the triangle singularity channels $3$ and $6$ improves the fits significantly. A further introduction of free parameters in the channels $5$, $7$ and $8$ seem to improve the fit, but do not improve the result per degree of freedom. Finally, introducing a pole which couples only to $\rho(770)\pi$ in a relative $S$-wave (channel $1$) improves the fits even further. As noted earlier, since the pole fit parameters $g_i$ and bare mass $m_\text{fit}$ are expected to be independent of the momentum transfer $t^\prime$, we consider an overall fit across the four different $t^\prime$ bins in which the $d_i$ are allowed to differ from bin to bin, but the $g_i$ and $m_\text{fit}$ are not. The results are given in Fig.~\ref{fig:full-fits}. The fit quality is noticeable improved towards lower energies. Indeed, below we will discuss that the pole could correspond to the $a_1(1260)$ resonance.

Introducing an additional coupling of the pole to channel $5$ did not improve the fit. An introduction of a second pole to test the pole contribution in the vicinity of the triangle singularity also did not improve the fit. The fits suggest that the triangle singularity arising from a rescattering is enough to sufficiently describe the data without the need of a genuine $a_1(1420)$. This has to be taken with a grain of salt as for example the order in which the parameters are turned on may have introduced a bias. Further, possible systematic uncertainties in the data can also lead to model misspecification through the simple $\chi^2$-minimization as discussed in Ref.~\cite{Sadasivan:2025kjj}. For alternative and more stable techniques based on neural networks see, e.g., Refs.~\cite{Sombillo:2020ccg, Sadasivan:2025kjj, Co:2024bfl}.

The fits with an integration over the outgoing two-body invariant mass given by Eq.~\eqref{eq:model-I} seem to provide better results than the fits with a fixed outgoing two-body invariant mass given by Eq.~\eqref{eq:model-F}. Additionally, we also provide the values of $\chi^2$ per degree of freedom values divided by a factor of $4$, called $\tilde{\chi}^2$ to account for the possibly significant systematic errors. While there is no physical motivation for this, this is very reasonable, since a large number of events are used in the experimental analyses, making the statistical errors negligible and the systematic errors the dominant source of errors.
\begin{figure}[t]
    \centering
    \includegraphics[width=0.8\linewidth]{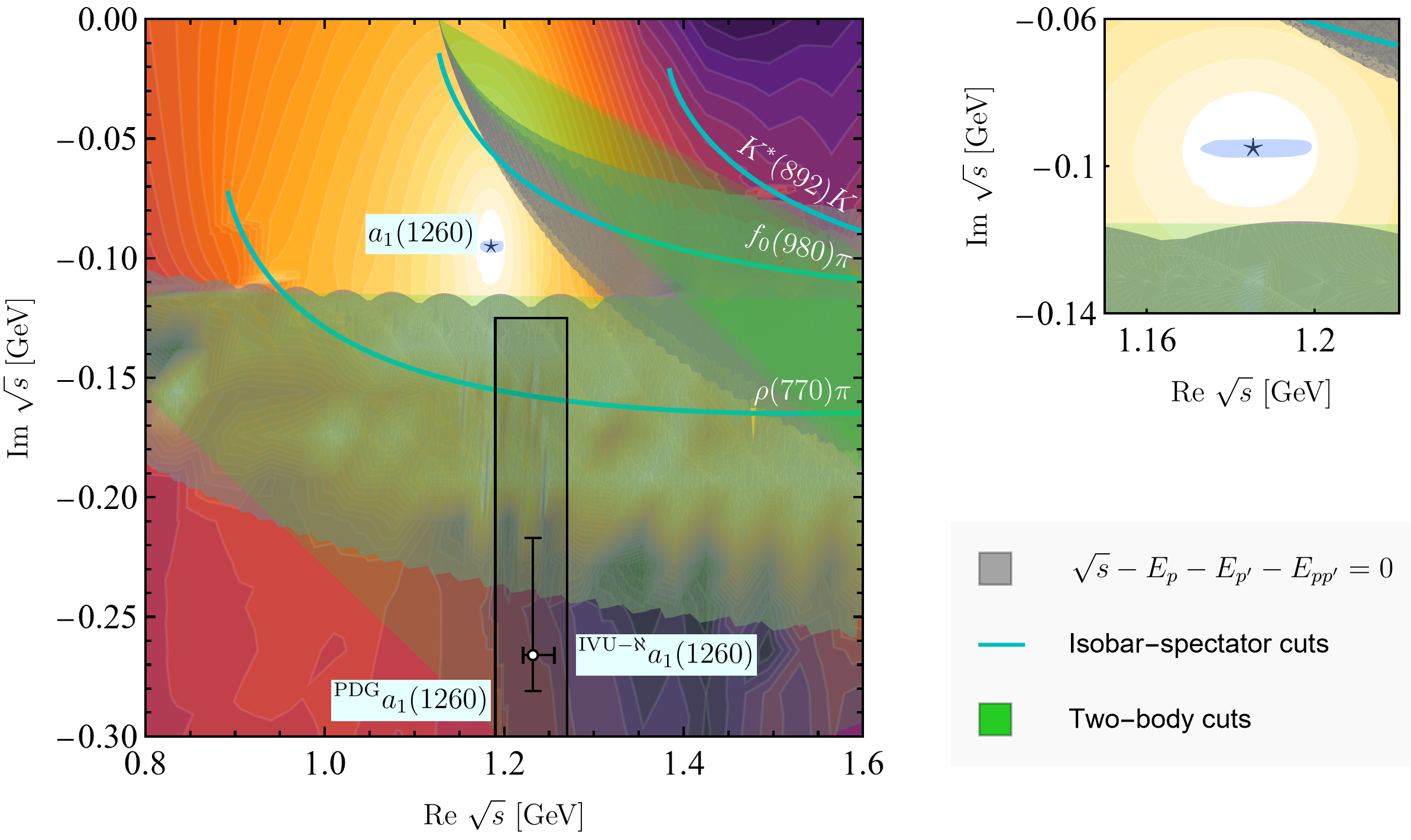}
    \caption{The $a_1(1260)$ pole in $T_{11}(\sqrt{s},p',p)$ of \cref{eq:LS-JLS} for fixed spectator momenta $p,\,p'$ in the fit I-5-1\textsubscript{1}G. The plot on the right shows the neighborhood of the pole zoomed in. The blue blob about the pole shows the uncertainty in the pole position induced by data uncertainties and obtained from samples taken around the $\chi^2$ minimum. The gray bands show the singularities arising in the IVU formalism -- these are exactly the singularities handled by contour deformation. They indicate ``forbidden'' areas in which no resonance poles can be searched for. The two-body amplitudes in the isobar component are evaluated on different sheets to expose the two-body resonances. This also involves rotating the standard two-body threshold cut by some angle~\cite{Doring:2025phq}. The cyan lines correspond to the isobar-spectator cuts associated with the complex thresholds from these two-body resonances. Note that the $f_0(500)\pi$ cut lies outside the plotted interval. The green bands correspond to the two-body threshold cuts in the various two-body amplitudes indicating other forbidden regions in which the SMC crosses a rotated two-body cut~\cite{Doring:2025phq}. The pole position of the $a_1(1260)$ from Ref.~\cite{Sadasivan:2021emk} and the PDG estimate~\cite{ParticleDataGroup:2020ssz} are also given.}
    \label{fig:a1pole}
\end{figure}%

In the fits given in Fig.~\ref{fig:full-fits}, a pole was included that couples to channel $1$, i.e., $\pi\rho$ in $S$-wave that has a known strong coupling to the $a_1(1260)$, with other channels coupling less to the $a_1$~\cite{Molina:2021awn}. Through the scattering equation, this bare pole acquires an imaginary part and moves into the complex $\sqrt{s}$ plane as shown for fixed spectator momenta in Fig.~\ref{fig:a1pole}. The figure shows not only the pole but also two-body complex thresholds and associated cuts from resonant isobars, as well as ``forbidden regions''. One source of these regions are three-body singularities at the zeros of \cref{eq:singularities1} and \cref{eq:singularities}. A second source is a geometric clash when the SMC integration contour of \cref{eq:SMC} crosses the two-body right-hand cuts of the isobar amplitudes, e.g., the $\pi\pi$ or $K\bar K$ cut. Pole searches in these regions are not allowed; they can be made accessible through cut rotations and SMC deformations. The details are described and illustrated in Ref.~\cite{Doring:2025phq}. In summary, a pole in the three-body amplitude can be isolated on the second Riemann sheet in a $\sqrt{s}$ region allowed by the chosen SMC and by the choice of the isobar two-body cuts.

While it is enticing to match this pole with a resonance with the same quantum numbers that couples to the same channel --the $a_1(1260)$-- it should be noted that, first of all, the experimental data also includes non-resonant contributions like the Deck effect. Furthermore, the data that are used for the fits are in a different channel than the one to which the pole couples predominantly. So the pole contribution does not appear at the leading order. However, we found the pole parameters to be rather stable in both cases when the outgoing two-body invariant mass $\sqrt{\sigma}$ is fixed or integrated over. In the former, the pole position is calculated as $\sqrt{s_{\rm pole}}=(1179^{+7}_{-10} - i\,93^{+1}_{-1})\MeV$, whereas in the latter, the pole position is calculated as $\sqrt{s_{\rm pole}}=(1185^{+15}_{-14} - i\,95^{+3}_{-2  })\MeV$. The quoted errors reflect only the provided statistical uncertainties on the input data, and we expect large uncertainties when systematic errors are included. We refrain here from calculating the residue function as in Ref.~\cite{Mai:2021nul} because the pole parameters are rather uncertain, anyway. Compared to the current PDG value of the pole position~\cite{ParticleDataGroup:2020ssz}, given by $\sqrt{s_{\rm pole}}=(1209^{+13}_{-10} - i288^{+45}_{-12})\MeV$, the imaginary part of the pole is substantially smaller. This is also the case when it is compared with the pole position obtained using the IVU formalism in a previous work~\cite{Sadasivan:2021emk}, given by $\sqrt{s_{\rm pole}}=(1232^{+15+9}_{-0-11} - i266^{+0+15}_{-22-27})\MeV$. However, the real part matches reasonably well. It is well-known that the non-resonant contributions to the amplitude shifts the pole position non-trivially. We still call this pole the $a_1(1260)$, although it should be taken with a grain of salt. If one wishes to investigate this further, it is better to use the experimental data in channel $1$. Nevertheless, it is remarkable how well the mass of the $a_1$ is predicted given that channel 5 is a sub-leading decay of that resonance~\cite{Molina:2021awn}, and the $a_1(1260)$ only manifest itself as a small shoulder at the lower energies of the triangle singularity structure.
\begin{figure}[t]
    \centering
    \includegraphics[width=0.45\linewidth]{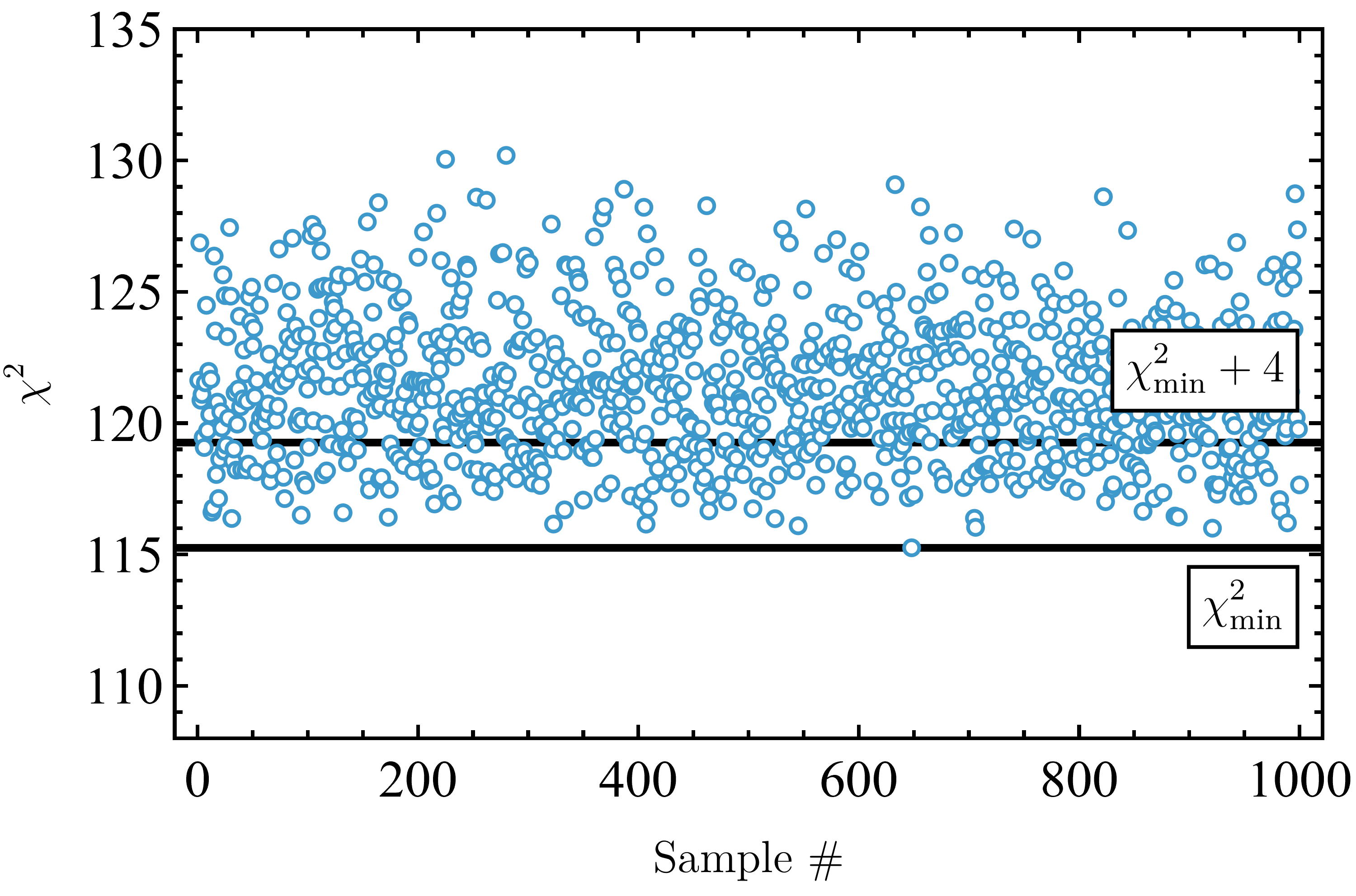}
    \caption{The distribution of $\chi^2$ values for samples about the parameters corresponding to $\chi^2_\textrm{min}$ corresponding to the fit I-5-1\textsubscript{1}G, see Tab.~\ref{tab:fit-values}. The interval containing $\chi^2$ values up to $\chi^2_\textrm{min} + 4$ is also given.}
\label{fig:chi2}
\end{figure}%

Due to large $\chi^2$ values, errors cannot be reliably determined and are in general underestimated. However, we try to give a basic estimate on the errors in order to test the stability of the formalism. The triangle singularity region is in general rife with problems, and a basic test of the stability of the formalism can give essential information. This also sets the stage for future analyses. For statistical errors, we did not perform sampling with replacement bootstrapping in order to avoid computational costs. Instead, we generated samples of order $10^3$ uniformly about the parameters corresponding to the minimum value of $\chi^2$, for the example of fit I-5-1$_1$ G. The distribution of the $\chi^2$ values are given in Fig.~\ref{fig:chi2}, which also shows a band containing the samples within $\chi^2_\textrm{min} + 4$ to see how the statistical errors propagate to the fits. In Fig.~\ref{fig:full-fits}, the variations in the amplitudes are given by the colored bands about the main curves corresponding to this choice. For numerical errors, we studied the effect of varying the SMC parameters about $10\%$ from the actual parameters, which, of course, shouldn't change the results at all, assuming analyticity and infinite numerical precision. One must be careful here since the amplitudes we are interested in are rich in singularities, and SMC parameters are chosen cautiously. The numerical errors are not given in Fig.~\ref{fig:full-fits}, however the total quadratic errors are given by the gray bands. The numerical errors are negligible and the total quadratic error is almost same as the statistical errors. The stability of the pole position can be seen in Fig.~\ref{fig:a1pole}, where the blue blob about the pole corresponds to the parameters within $\chi^2_\textrm{min} + 4$.

\section{Summary and Conclusions}
Triangle singularities (TS) are not genuine resonances but kinematic enhancements originating from three hadrons being on-shell. Traditionally, these structures have been calculated by considering one-loop Feynman diagrams. However, that contribution can be understood as the approximation of a unitary three-body scattering amplitude. As such, this formalism also needs to have at least two channels, because the TS involving heavier particles is usually visible in the lineshapes of lighter decay products. 

In the present study, previous, semi-quantitative work on one of the most prominent TS, the $a_1(1420)$ has been extended to a coupled-channel framework involving all possible combinations of pions and kaons (9 channels), all isobars up to $P$-wave, and sub-amplitudes matched to experimental two-body phase shifts and inelasticities. In addition, the formalism contains coupled channels in the isobars themselves, like the $\pi\pi\to K\bar K$ transition. A qualitative analysis reveals that additional channels and rescattering can lead to moderate changes in the maximum of the TS in $\sqrt{s}$ by some 20 MeV which is less than the width of the structure. While these effects beyond the one-loop approximation can substantially change the overall scale of the TS, that scale is unobservable in principle. A new observation is that TS can manifest themselves as dips instead of enhancements, depending on the size and sign of the direct production of the isobar-spectator pair without any rescattering. Here, the data shows an enhancement. 

To quantitatively study the effect of the triangle singularity, the production amplitude is fitted to COMPASS data available for four different bins in momentum transfer $t'$. These bins can be connected in global fits. While data are qualitatively reproduced, the tiny statistical errors still lead to large overall $\chi^2$, probably due to systematic uncertainties, and a stability analysis of the results is performed. In particular, introducing a bare $s$-channel singularity leads to a substantial improvement of the fit, with a pole corresponding not to the TS, but to the nearby $a_1(1260)$ resonance. The signal is remarkably stable and the resonance pole, obtained through analytic continuation of the three-body amplitude, is remarkably close to the PDG region for this resonance. 

In summary, it has been demonstrated that large, 9-channel, unitary three-body amplitudes can be fitted to experimental lineshapes including the analytic continuation to extract resonance poles. Different $t'$ bins can be connected in a global analysis. Here, the analysis served to quantitatively study the $a_1(1420)$ TS. Coupled channels and rescattering effects lead to moderate changes of lineshapes, and the fits overall confirm the TS hypothesis. In addition, there is evidence for the $a_1(1260)$ being present in the amplitude. The current study paves the way to future analyses including simultaneously data from multiple final states and from Dalitz plots, revealing additional kinematic content of the intricate multi-channel interplay of resonances and triangle singularities. The code used in this work is also made available under the arXiv ID of this article on GitHub~\cite{Sakthivasan_source-code_2026}.

\section*{Acknowledgments}
We thank Bernhard Ketzer, Mikhail Mikhasenko, David Spülbeck, Annika Thiel, and other members of the COMPASS group for useful comments and providing the results of their analysis. 
We also thank Feng-Kun Guo for a critical reading of the manuscript.
The work of AS was funded in part by the Ministry of Culture and Science of North Rhine-Westphalia through the NRW-FAIR project.
The work of YF and MD was supported by the National Science Foundation (NSF) Grant No. 2310036. 
The work of MM was further funded through the Heisenberg Programme by the Deutsche Forschungsgemeinschaft (DFG, German Research Foundation), No. 532635001.
The work of MD is also supported by the U.S. Department of Energy,
Office of Science, Office of Nuclear Physics under SURATech LLC/DOE contract 89243126CSC000213 (previously contract DE-AC05-06OR23177).

\bibliography{BIB}
\clearpage
\appendix
\section{Experimental Data}
\label{app:exp-data}
In this section, we give the experimental dataset that was used for the fits in this work. Since the actual dataset was not accessible to us, the following was extracted from the plots given in Ref.~\cite{COMPASS:2015gxz}. There are four different values of momenta transferred between the proton and the pion ($t^\prime$) considered in the analysis: $0.100<t^\prime/\textrm{GeV}<0.141$ (see Tab.~\ref{tab:intensities-t1}), $0.141<t^\prime/\textrm{GeV}<0.194$ (see Tab.~\ref{tab:intensities-t2}), $0.194<t^\prime/\textrm{GeV}<0.326$ (see Tab.~\ref{tab:intensities-t3}) and $0.326<t^\prime/\textrm{GeV}<1.000$ (see Tab.~\ref{tab:intensities-t4}). All data correspond to $0.96<m_{\pi^-\pi^+}/\textrm{GeV}<1.00$, that is, the outgoing isobar invariant mass interval is taken to contain the $f_0(980)$.
\begin{table}[t]
    \setlength{\tabcolsep}{8pt}
    \centering
    \begin{tabular}{c c c}
         $m_{3\pi}~[\textrm{GeV}]$ & (Intensity$/40~\textrm{MeV})/10^3$ & (Error$/40~\textrm{MeV})/10^3$ \\ \hline
         1.16 & 1.28 & 0.13 \\
         1.20 & 2.69 & 0.19 \\
         1.24 & 5.15 & 0.22 \\
         1.28 & 6.73 & 0.24 \\
         1.32 & 7.51 & 0.24 \\
         1.36 & 8.72 & 0.26 \\
         1.40 & 9.39 & 0.24 \\
         1.44 & 5.56 & 0.2 \\
         1.48 & 2.97 & 0.14 \\
         1.52 & 1.92 & 0.11 \\
         1.56 & 1.28 & 0.09 \\
         1.60 & 1.00 & 0.07 \\
    \end{tabular}
    \caption{Intensities for $0.100<t^\prime/\textrm{GeV}^2<0.141$ with $0.96<m_{\pi^-\pi^+}/\textrm{GeV}<1.00$.}
    \label{tab:intensities-t1}
\end{table}%
\begin{table}[t]
    \setlength{\tabcolsep}{8pt}
    \centering
    \begin{tabular}{c c c}
         $m_{3\pi}~[\textrm{GeV}]$ & (Intensity$/40~\textrm{MeV})/10^3$ & (Error$/40~\textrm{MeV})/10^3$ \\ \hline
         1.16 & 1.39 & 0.17 \\
         1.20 & 1.69 & 0.15 \\
         1.24 & 3.70 & 0.19 \\
         1.28 & 4.30 & 0.18 \\
         1.32 & 6.50 & 0.22 \\
         1.36 & 6.12 & 0.2 \\
         1.40 & 7.04 & 0.22 \\
         1.44 & 4.55 & 0.17 \\
         1.48 & 2.58 & 0.13 \\
         1.52 & 2.01 & 0.11 \\
         1.56 & 1.22 & 0.08 \\
         1.60 & 1.14 & 0.08 \\
    \end{tabular}
    \caption{Intensities for $0.141<t^\prime/\textrm{GeV}^2<0.194$ with $0.96<m_{\pi^-\pi^+}/\textrm{GeV}<1.00$.}
    \label{tab:intensities-t2}
\end{table}%
\begin{table}[t]
    \setlength{\tabcolsep}{8pt}
    \centering
    \begin{tabular}{c c c}
         $m_{3\pi}~[\textrm{GeV}]$ & (Intensity$/40~\textrm{MeV})/10^3$ & (Error$/40~\textrm{MeV})/10^3$ \\ \hline
         1.16 & 0.73 & 0.10 \\
         1.20 & 1.58 & 0.14 \\
         1.24 & 3.21 & 0.17 \\
         1.28 & 4.44 & 0.17 \\
         1.32 & 5.91 & 0.2 \\
         1.36 & 7.55 & 0.21 \\
         1.40 & 8.35 & 0.22 \\
         1.44 & 5.73 & 0.19 \\
         1.48 & 3.58 & 0.15 \\
         1.52 & 2.37 & 0.12 \\
         1.56 & 2.18 & 0.09 \\
         1.60 & 1.79 & 0.10 \\
    \end{tabular}
    \caption{Intensities for $0.194<t^\prime/\textrm{GeV}^2<0.326$ with $0.96<m_{\pi^-\pi^+}/\textrm{GeV}<1.00$.}
    \label{tab:intensities-t3}
\end{table}%
\begin{table}[t]
    \setlength{\tabcolsep}{8pt}
    \centering
    \begin{tabular}{c c c}
         $m_{3\pi}~[\textrm{GeV}]$ & (Intensity$/40~\textrm{MeV})/10^3$ & (Error$/40~\textrm{MeV})/10^3$ \\ \hline
         1.16 & 0.18 & 0.05 \\
         1.20 & 0.49 & 0.08 \\
         1.24 & 1.27 & 0.12 \\
         1.28 & 2.13 & 0.11 \\
         1.32 & 2.96 & 0.13 \\
         1.36 & 4.03 & 0.15 \\
         1.40 & 4.94 & 0.17 \\
         1.44 & 3.75 & 0.14 \\
         1.48 & 2.76 & 0.12 \\
         1.52 & 2.55 & 0.11 \\
         1.56 & 1.87 & 0.09 \\
         1.60 & 1.69 & 0.1 \\
    \end{tabular}
    \caption{Intensities for $0.326<t^\prime/\textrm{GeV}^2<1.000$ with $0.96<m_{\pi^-\pi^+}/\textrm{GeV}<1.00$.}
    \label{tab:intensities-t4}
\end{table}

\section{Polarization Vectors}
\label{app:pol}
The vertices given by Eq.~\eqref{eq:v feng2024} that enter the equation for the $B$-term as given by Eq.~\eqref{eq:B-term} contains the polarization vectors for isobars with a non-zero total spin. We work in the reference frame where the incoming isobar momenta are along the $z$-axis, whereas the outgoing ones lie on the $x-z$-plane with an angle $\theta$ about the $z$-axis. The polarization vectors for the incoming isobars read
\begin{equation}
    \begin{aligned}
        \epsilon_{-1}^\prime(\bm{p}) = \frac{1}{\sqrt{2}} \begin{pmatrix}
            0 \\ 1 \\ i \\ 0
        \end{pmatrix}, &&
        \epsilon_{0}^\prime(\bm{p}) = \frac{1}{m_\text{isobar}} \begin{pmatrix}
            \vert \bm{p} \vert \\ 0 \\ 0 \\ -\sqrt{m_\text{isobar}^2 + \bm{p}^2}
        \end{pmatrix}, &&
        \epsilon_{+1}^\prime(\bm{p}) = \frac{1}{\sqrt{2}} \begin{pmatrix}
            0 \\ -1 \\ i \\ 0
        \end{pmatrix},
    \end{aligned}
\end{equation}
where $\bm{p}$ is the magnitude of the three-momentum, and $m_\text{isobar}$ is the central value of the real mass as given in PDG of the relevant isobar. The subscripts $\pm1$, $0$ denote the helicity of the relevant isobar. We consider only massive spin-$1$ isobars in our work. The polarization vectors for the outgoing isobars read
\begin{equation}
    \begin{aligned}
        \epsilon_{-1}(\bm{p}) = \frac{1}{\sqrt{2}} \begin{pmatrix}
            0 \\ \cos\theta \\ i \\ -\sin\theta
        \end{pmatrix}, &&
        \epsilon_{0}(\bm{p}) = \frac{1}{m_\text{isobar}} \begin{pmatrix}
            \vert \bm{p} \vert \\ -\sqrt{m_\text{isobar}^2 + \bm{p}^2} \sin \theta \\ 0 \\ -\sqrt{m_\text{isobar}^2 + \bm{p}^2} \cos\theta
        \end{pmatrix}, &&
        \epsilon_{+1}(\bm{p}) = \frac{1}{\sqrt{2}} \begin{pmatrix}
            0 \\ -\cos\theta \\ i  \\ \sin\theta,
        \end{pmatrix}
    \end{aligned}
\end{equation}
where the quantities are the same as before.
\end{document}